\documentclass[a4paper,12pt]{article}
\usepackage{amsfonts}
\usepackage{graphicx}
\usepackage{amsmath} 
\graphicspath{ {Images/} }
\usepackage{french}
\usepackage{enumitem}
\usepackage{pifont}
\usepackage{geometry}
\usepackage[T1]{fontenc}
\usepackage{graphicx}
\usepackage{amsthm}
\usepackage{amsfonts}
\usepackage{pifont}
\usepackage{numprint}
\usepackage{hyperref}
\usepackage{siunitx}
\usepackage{textcomp}
\usepackage{multicol}
\usepackage{fancyhdr}
\usepackage{csquotes}
\usepackage{chronosys}
\usepackage{mathtools}
\usepackage{algorithm}
\usepackage{todonotes}
\usepackage{wrapfig}
\usepackage{enumitem}
\usepackage{tcolorbox}
\usepackage{hyperref}
\usepackage{envmath}
\usepackage{cases}
\usepackage{empheq}
\usepackage{array}
\usepackage{pdfpages}
\usepackage{minitoc}
\usepackage{multirow}
\usepackage{caption}
\usepackage{subcaption}

\usepackage{csquotes}

\usepackage{cite}

\graphicspath{ {figures/} }

\hypersetup{
    colorlinks=true,
    linkcolor=blue,
    filecolor=magenta,      
    urlcolor=blue,
    pdfpagemode=FullScreen,
    citecolor = blue
    }
\begin{document}
\begin{titlepage}
\newcommand{\HRule}{\rule{\linewidth}{0.5mm}}
\center
\textsc{\LARGE
Rapport de Stage de Master
} \\[1cm]
\includegraphics[width=90mm, scale=1.5]{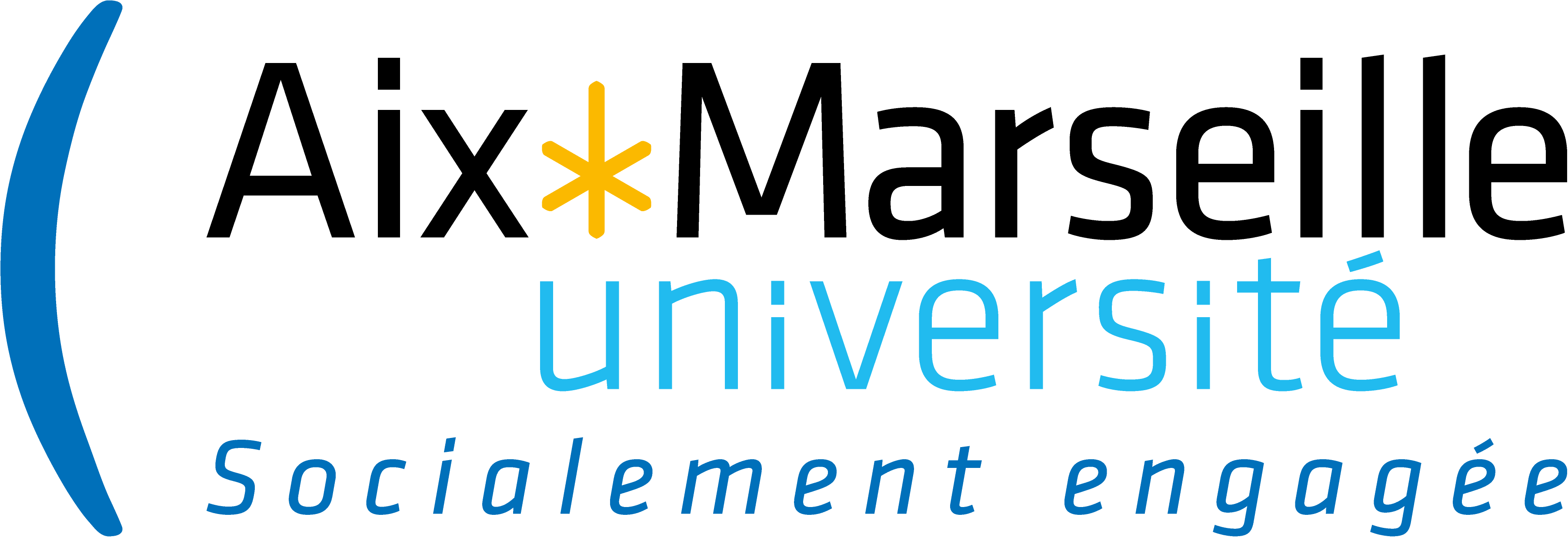} \\[1cm]
\includegraphics[width=90mm, scale=1.5]{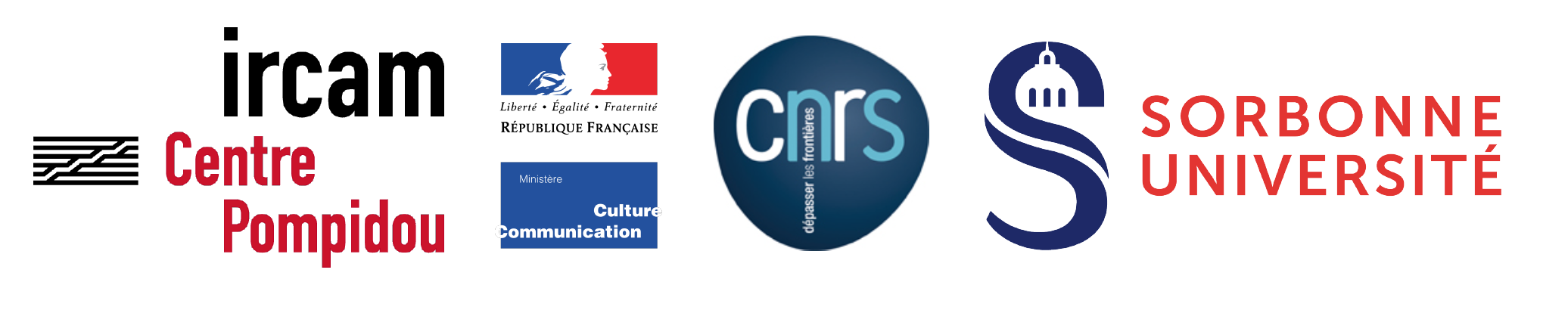}\hfill
\includegraphics[width=90mm, scale=1.5]{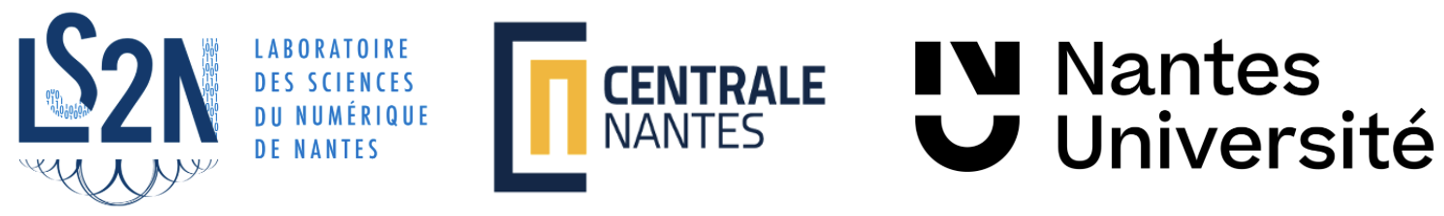}\\[1cm]
\HRule \\[0.1cm]
{ \huge \bfseries Réduire le bruit grâce à la Réalité Augmentée Sonore - \textit{Auditory Concealer} \\[0.15cm] }
\HRule \\[1.5cm]
{\Large Clara \textsc{boukhemia}
\\[1.5cm]
Encadrants : Nicolas \textsc{misdariis} (STMS, IRCAM) \\ \& Mathieu \textsc{lagrange} (LS2N, CNRS) \\
Responsable pédagogique : Antoine \textsc{gonot} \\
Rapporteur : Mitsuko \textsc{aramaki}
\\[1.5cm]\today
\\ [1.5cm]}
{\LARGE M2 Acoustique \& Musicologie}
\\{\Large Année Universitaire 2023--2024}
\end{titlepage}

\clearpage
{ \huge \bfseries Résumé}
\\[1cm]

Ce rapport présente le travail de $22$ semaines de stage au sein de l'équipe Perception et Design Sonores du laboratoire Sciences et Technologies de la Musique et du Son (STMS) à l'Institut de Recherche et de Coordination Acoustique/Musique (IRCAM). Au lancement du projet \textit{Reducing Noise with Augmented Reality (ReNAR)}, qui vise à créer un outil permettant de réduire en temps-réel l'impact cognitif des sons perçus comme désagréables ou gênants au sein d'environnements intérieurs, une première étude a été réalisée dans le but de valider la faisabilité et l'efficacité d'une nouvelle approche de masquage nommée \textit{concealer}. L'hypothèse principale réside dans la possibilité que l'approche \textit{concealer} fournisse de meilleurs résultats qu'une approche dite \textit{masker} en termes d'agrément perçu. Des mixtures entre deux sources de bruit (ventilation) et cinq maquilleurs (sons d'eau) ont été générées à l'aide des deux approches selon plusieurs niveaux. L'évaluation de l'agrément perçu de ces mixtures a montré que l'approche \textit{masker} reste meilleure que l'approche \textit{concealer}, quelque soit la source de bruit, le son d'eau, ou le niveau utilisé. \\

\textbf{Mots-clefs} : \textit{approche concealer, approche masker, agrément, bruit de ventilation, sons d'eau}

\clearpage
{ \huge \bfseries Remerciements}
\\[1cm]

Je tiens en premier lieu à remercier mes deux encadrants, pour leur bienveillance, leur confiance et leurs précieux conseils tout au long de ce stage. 

Merci à Nicolas Misdariis pour m'avoir formée au développement d'une expérience perceptive. Son insatiable curiosité, son optimisme et sa capacité à toujours s'émerveiller des nouvelles propositions artistiques me serviront d'exemple pour la suite de mon projet professionnel.

Merci à Mathieu Lagrange pour son aide et ses explications sur les aspects théoriques et pratiques du traitement du signal. Son ouverture d'esprit et son expérience de chercheur m'ont donné confiance en moi pour continuer dans cette voie. \\

Je remercie également toute l'équipe Perception et Design Sonores, qui m'ont accueillie dans une ambiance joviale, et qui ont fait de ce stage une période plus qu'agréable. Un grand merci pour ces pauses de midi soldées de points cultures cinématographiques, littéraires, et évidemment musicaux ! Je remercie particulièrement les doctorants, Matthieu, Michèle et Armand, mes voisins de bureau, pour leur aide et leur humour. \\

Comment ne pas dédier quelques mots à mes incroyables parents, pour les remercier de leur soutien sans faille. Merci de toujours croire en moi et d'être venu apporter un peu du soleil marseillais à Paris ! \\

Un grand merci à mon super-copain, pour sa gentillesse sans limite et son soutien à toute épreuve. Merci d'être là. \\

Enfin, pour clore un chapitre et en ouvrir un nouveau, merci à toute l'équipe pédagogique du master Acoustique et Musicologie, et en particulier à Antoine Gonot, pour son écoute et son énergie. Merci à mes camarades et amis, avec qui j'ai passé deux années riches en joie et en rigolade ! 

\clearpage
\tableofcontents

\newpage
\listoffigures

\newpage
\section{Introduction}
Défini par le Ministère de la Transition Écologique et de la Cohésion des Territoires comme \textit{"un phénomène acoustique produisant une sensation auditive considérée comme désagréable ou gênante"}, le bruit représente un problème très présent dans le quotidien de beaucoup de personnes. Parfois constamment soumises à l'excès de bruit, ces personnes peuvent voir leur santé se détériorer d'un point de vue physiologique (endommagement du système auditif entraînant des lésions ou des pertes auditives), mais aussi d'un point de vue psychologique (apparition de symptômes tels que la perturbation du sommeil ou le stress). Ces nuisances sonores, résultat de la pollution sonore rencontrée dans plusieurs types d'environnements, soulèvent donc de réels enjeux de santé publique.

Pour tenter d'améliorer le confort des personnes au sein de leurs espaces de vie, le projet \textit{Reducing Noise with Augmented Reality (ReNAR)} financé par l'Agence Nationale de la Recherche (ANR) propose de créer un outil permettant de réduire en temps-réel l'impact cognitif des sons perçus comme désagréables ou gênants au sein d'environnements intérieurs. Pour cela, deux approches de réalité augmentée sonore sont étudiées en parallèle au sein du laboratoire STMS de l'IRCAM à Paris, du LS2N à Nantes et du Loria à Nancy, grâce à une collaboration entre les chercheurs Nicolas MISDARIIS (Paris), Mathieu LAGRANGE (Nantes) et Romain SERIZEL (Nancy). Le travail autour des deux approches est mené selon deux axes. Le premier, intitulé \textit{Auditory Concealer}, a pour but de déterminer quels sont les aspects importants de l’agrément ou du confort perçu des scènes acoustiques, et d'identifier les sources sonores gênantes ou désagréables présentes dans les environnements étudiés. Le deuxième, intitulé \textit{Speech Intelligibility Reducer}, s'intéresse plus spécifiquement à la parole intelligible et vise à réduire l'impact négatif des conversations environnantes sur les personnes partageant un espace. Pour les deux axes, l'approche développée consiste à ajouter un signal à une scène sonore contenant un son indésirable, tout en minimisant l'augmentation du niveau sonore global de cette scène.

Le travail effectué dans le cadre du stage constitue un point d'entrée dans le projet \textit{ReNAR} pour l'axe \textit{Auditory Concealer}, avec une première étape dans le développement de l'approche \textit{concealer} qui se focalise sur un type d'environnement bien précis : les bureaux ouverts. Dans la catégorie des environnements intérieurs, ce type d'espace de travail n'échappe pas au problème du bruit, et ce depuis sa conception dans les années 1950. Les nuisances sonores se traduisent dans ce cas par la présence de sources sonores saillantes, dérangeantes et indésirables qui peuvent polluer le paysage sonore global et le rendre désagréable, voire ennuyeux pour les personnes qui y travaillent. Partant du cas d'étude d'un système de ventilation, un questionnement s'est posé sur la faisabilité et l'efficacité de l'approche \textit{concealer} face à une approche de masquage nommée approche \textit{masker}, souvent retrouvée dans la littérature. En particulier, il est intéressant de se demander si l'approche \textit{concealer} se trouve être plus efficace que l'approche \textit{masker} en termes d'agrément, et ce pour un niveau global proche de celui du bruit initial. Pour tenter de répondre à cette question, une revue de l'état de l'art sera tout d'abord explicitée en section \ref{etat-art}. La construction de l'approche \textit{concealer} sera ensuite détaillée en section \ref{section3}, pour enfin la comparer en section \ref{expe} à l'approche \textit{masker} à l'aide d'une expérience perceptive permettant d'évaluer l'agrément perçu. Cette dernière section présentera les premiers résultats de cette expérience, et une amorce de discussion conclura ce rapport pour ouvrir sur la suite du projet.

\newpage
\section{État de l'art}\label{etat-art}
Avant d'entamer une réflexion en lien avec le projet \textit{ReNAR}, cette partie reprendra les notions importantes citées dans l'état de l'art complet réalisé au début de la période de stage afin de poser un cadre général sur ces notions et la problématique du bruit.

\subsection{Qu'est-ce que le confort d'un point de vue sonore ?}\label{confort-sonore}
\begin{figure}[h!]
    \centering
    \includegraphics[scale = 0.5]{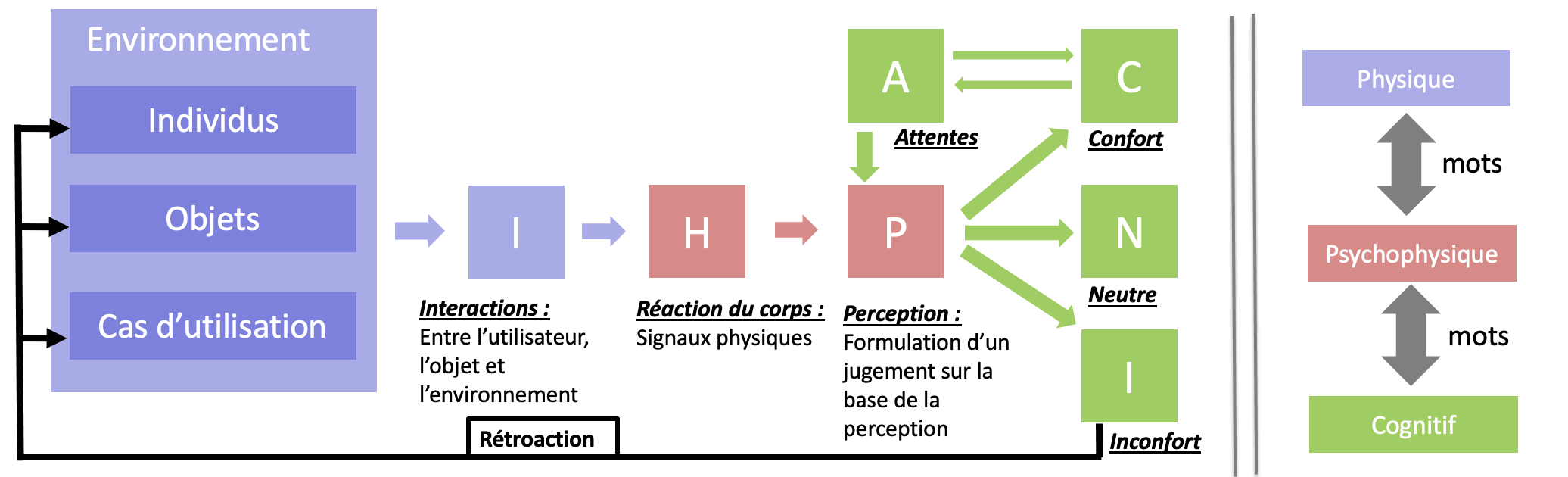}
    \caption{Modèle de confort proposé par Vink et Hallbeck \cite{vink_editorial_2012}, repris et traduit par Matthieu Duroyon \cite{duroyon2025confort}}
    \label{modele-confort}
\end{figure}

Pour parler de confort en utilisant les termes "Confort/Inconfort", il est important d'introduire une certaine contextualisation. En effet, les modèles de confort présentés dans la littérature établissent un lien étroit entre objet, environnement et utilisateur. Celui que proposent Vink et Hallbeck \cite{vink_editorial_2012}, et qui est repris par Matthieu Duroyon dans le cadre de sa thèse \cite{duroyon2025confort} permet de bien se rendre compte de ces interactions (Figure \ref{modele-confort}). Comme l'expliquent Duroyon et coll. par leur définition, \textit{“le confort est une sensation subjective rendant un objet manufacturé agréable dans son utilisation et cohérent avec les attentes des utilisateurs”} \cite{duroyon_how_2024}. Les attentes et la satisfaction des personnes concernant un produit entrent ainsi en compte dans l'évaluation du confort. 

Mais de quel confort parle-t-on ? S'il est clair que celui-ci concerne le son, il est possible de faire la différence entre plusieurs types de conforts. Soucieux d'utiliser un vocabulaire adapté lors de sa recherche sur le confort au sein des véhicules électriques, Duroyon en différencie deux types liés au son \cite{duroyon2025confort}. D'une part le confort "acoustique”, qui s’approche d’un travail sur la “quantité sonore”, visant notamment à la réduire. D'autre part, le confort "sonore”, qui s’accorde davantage avec la “qualité sonore”, et suggère de s’intéresser à la perception des environnements. \\

Par ailleurs, le jugement d'un produit manufacturé impose un cadre industriel à l'évaluation du confort, qu'il soit acoustique ou sonore, mais il est possible de ramener ce jugement à des environnements entiers. Pour parler du confort à bord des trains ou des véhicules électriques par exemple, les adjectifs "Confortable" et "Inconfortable" sont utilisés pour décrire l'environnement comme un produit dont les personnes se servent au quotidien. Les études visant à évaluer le confort dans de tels environnements se déroulent la plupart du temps \textit{in situ}, afin de mettre les usagers en conditions réelles, et de leur fournir un contexte qui se veut le plus réaliste possible. C'est ce que font notamment Mzali et coll. en 2000 \cite{mzali_auditory_2000} et Kossachka et coll. en 2006 \cite{kossachka_approche_2006} lors de leurs études sur le confort à bord des trains, utilisant une approche psycholinguistique. Il est également intéressant de citer la méthode d'écoute réactivée, explicitée par Augoyard en 2001 dans son article \textit{L'entretien sur écoute réactivée} \cite{augoyard_entretien_2001}, dont le but est de laisser les habitués d'un lieu expérimenter les séquences sonores typiques de ce lieu avant de mener un entretien pour récolter leurs ressentis et interprétations de l'environnement sonore. 

\subsection{Étudier la gêne pour améliorer le confort}
En définissant le confort, il est apparu qu'une mesure ou une évaluation de ce dernier demanderait une contextualisation poussée, obtenue par la mise en place d'une expérience \textit{in situ}. Une telle expérience étant difficile à mettre en œuvre, il est possible de s'intéresser à des notions étroitement liées à la notion de confort (si ces liaisons ne sont pas explicitement étudiées, elles paraissent assez intuitives dans la littérature qui traite du confort). C'est le cas de la notion de gêne, qui est plus étudiée dans la littérature et directement applicable au son grâce aux adjectifs "Gênant" et "Non gênant" (ce qui n'est pas le cas du confort, puisqu'un son est rarement évalué comme "confortable" ou "inconfortable"). \\

Plusieurs protocoles sont utilisés pour mesurer la gêne, qui permettent de récolter des données objectives et subjectives, mais qui semblent également requérir une certaine contextualisation. En effet, la notion de gêne est souvent liée à une action ou une tâche, que celle-ci soit réelle ou imaginée. Selon Guski et coll. \cite{guski_concept_1999}, \textit{"la gêne constitue une réaction à la perturbation de l'activité causée par des événements bruyants"}. D'après les conclusions de Misdariis et coll. \cite{misdariis_community_2019} et de Marquis-Favre et coll. \cite{marquis-favre_laboratory_2019} lors de leurs études sur la gêne perçue, il est alors possible de simplement demander à des personnes d'effectuer, ou de s'imaginer effectuer une tâche pour évaluer la gêne, que ce soit lors d'une expérience \textit{in situ} ou en laboratoire. Il est important de préciser gêne "perçue", car celle-ci est évaluée de manière subjective, en demandant aux personnes qui effectuent la tâche de décrire à quel point ils se sentent gênés. Une autre mesure, cette fois objective, peut être envisagée en récoltant la performance à la tâche demandée, celle-ci donnant une indication sur la gêne induite par les sons. En étudiant l'évolution de cette performance, il est possible d'évaluer une gêne autre que perçue (une décroissance de performance indiquant un niveau de gêne élevé). C'est ce que mettent en place Brocolini et coll. \cite{brocolini_effect_2016} 
lors de leur étude visant à évaluer la gêne induite par la présence de différentes sources sonores dans les bureaux ouverts. \\

Finalement, s'il semble aller de soi qu'une réduction de la gêne due aux sons présents dans un environnement permettrait d'en améliorer le confort, quelques points posent tout de même question. Tout d'abord, puisque la gêne s'évalue en relation avec une action, il est nécessaire comme pour le confort de poser certains éléments de contexte. Aussi, il semble apparaître deux types de gêne : d'une part la gêne dite "spécifique", ressentie par l'auditeur lorsque celui-ci est soumis au bruit d'une unique source sonore. Ce premier type de gêne a notamment été étudié par Sporer et coll. \cite{sporer_distracting_2016} dans le but d'identifier les sources sonores jugées les plus gênantes. D'autre part, la gêne dite "totale", ressentie par l'auditeur lorsque celui-ci est soumis au bruit de plusieurs sources sonores combinées, pourrait permettre de se rapprocher un peu plus de la réalité d'une scène sonore complexe. Le lien entre gêne spécifique et gêne totale n'étant pas évident (la somme des gênes spécifiques ne permet pas de représenter la gêne totale a priori), il semble difficile d'imaginer se faire une idée de la gêne totale induite par l'ambiance sonore d'un environnement composé de différentes sources sonores en mesurant la gêne spécifique induite par chacune de ces sources sonores. 

\subsection{Étudier l'agrément pour améliorer le confort}
La définition d'une tâche à réaliser étant nécessaire pour évaluer la gêne, il est possible de se tourner vers une dernière notion pour simplifier l'évaluation d'ambiances sonores : l'agrément. Désignée par les adjectifs "Plaisant" et "Non plaisant", ou encore "Agréable" et "Désagréable", un lien intuitif peut être établi entre cette notion et celle de confort. En effet, il est peu probable qu'un environnement sonore confortable soit désagréable, et la notion d'agrément apparaît clairement dans la définition de Duroyon et coll. \cite{duroyon_how_2024} citée en section \ref{confort-sonore}. Une première réflexion tend à dire que l'agrément est une notion subjective relevant des goûts de chacun. C'est une sensation qui vient naturellement et qui n'est pas forcément explicable. Étudiée à l'aide de mesures subjectives d'agrément "perçu", il est le plus souvent demandé aux participants d'expériences perceptives de juger s'ils trouvent un son ou un environnement sonore agréable ou désagréable, sans parler de tâche à réaliser comme pour la gêne. \\

Plusieurs études se sont intéressées à l'agrément d'environnements sonores, et l'évaluation de cette notion à permis de la mettre en relation avec certains facteurs. Par exemple, l'étude de Liebetrau et coll. a permis de faire des liens entre certains facteurs acoustiques comme la rugosité et la netteté et l'agrément perçu \cite{liebetrau_quantifying_2015}. Des études de corrélations semblent en effet montrer que plus un son est net et rugueux, moins il est plaisant. D'autre part, l'étude de Guillén et coll., qui s'inscrit dans le contexte de la perception sonore en milieu urbain, met en avant un certain nombre de facteurs non-acoustiques apparemment corrélés avec la notion d'agrément \cite{domingo_importance_2007}. Parmi eux se trouvent notamment l'aspect visuel des environnements et le comportement des participants vis-à-vis du bruit. \\

Cela étant dit, il est important de préciser que l'évaluation de l'agrément peut se faire de manière globale ou de manière continue. L'évaluation globale se fait généralement à l'aide d'une échelle sémantique, à la manière de Guillén et coll. \cite{domingo_importance_2007}, qui présentent aux participants une échelle de jugement allant de 1 pour \textit{"Very Unpleasant"} à 7 pour \textit{"Very pleasant"}. Ceux-ci expérimentent d'abord des séquences sonores, puis utilisent l'échelle pour les juger en termes d'agrément. L'évaluation continue, quant à elle, peut aussi se faire à l'aide d'une échelle sémantique, utilisée cette fois par les participants pendant l'écoute des sons ou des scènes sonores. Cette seconde méthode a notamment été utilisée par Aumond et coll. lors de leur étude visant à évaluer l’agrément général d’un environnement sonore le long d’un trajet urbain, qui montre que l’agrément ressenti global donné par les participants après avoir été immergés dans une scène sonore est principalement expliqué par la moyenne de toutes les valeurs recueillies \cite{aumond_global_2017}. Baptiste Bouvier utilise également une évaluation continue dans le cadre de sa thèse, qui traite de la saillance auditive dans le cadre de la perception de l'environnement sonore \cite{bouvier_saillance_2024}. Cette évaluation concerne néanmoins le désagrément, et non l'agrément perçu. Le but étant d'étudier le potentiel lien entre saillance et désagrément, il justifie sa volonté de ne pas se restreindre à une mesure globale d'abord par souci de comparaison. En effet, la saillance a été mesurée de manière continue (une scène sonore peut contenir différents évènements sonores saillants), il faut donc une mesure du même type pour le désagrément pour comparer les deux notions. Aussi, dans un cas où seul le désagrément serait étudié, Bouvier affirme qu'une restitution globale rétrospective est affectée par différents effets, notamment l'effet de récence et de primauté, décrit par Aumond et coll. \cite{aumond_global_2017} comme l'effet \textit{"par lequel les jugements initiaux et finaux d’une séquence sont mieux mémorisés au moment où l’évaluation rétrospective est donnée"}. Il est donc selon lui plus souhaitable d'étudier le désagrément en continu pour se rapprocher de l'expérience des usagers immergés dans un paysage sonore. \\

Concernant l'agrément en tant que notion, regroupant à la fois ses aspects positif et négatif (agrément et désagrément), aucune définition ne semble donnée par la littérature. Les recherches sur l'agrément perçu dans un contexte sonore donnent néanmoins matière à réflexion sur les méthodes d'évaluation et les types de facteurs responsables de cette sensation subjective. Parmi ces derniers, des indicateurs acoustiques peuvent être évalués grâce à des corpus sonores représentatifs du paramètre étudié, et des indicateurs non-acoustiques peuvent être recueillis à l'aide de questionnaires. De plus, même si l'agrément dépend des goûts de chacun, il est apparu dans la littérature que certains sons sont communément acceptés, c'est-à-dire qu'ils paraissent agréables à l'oreille humaine en général. C'est par exemple le cas des sons naturels. L'étude de Cai et coll., qui vise à évaluer l'effet du masquage faisant intervenir des sons d'eau sur la perception du bruit industriel, semble confirmer que les sons d'eau sont jugés de manière positive \cite{cai_effect_2019}. Il serait alors possible de dresser une typologie des sons agréables et désagréables, aidant à améliorer l'agrément perçu, et possiblement le confort par extension.

\subsection{Le masquage pour traiter le bruit au niveau perceptif}
Les mesures de gêne ou d'agrément représentent de bonnes méthodes pour faire un état des lieux de la perception d'environnements sonores, mais elles ne permettent pas d'en améliorer le confort. Dans le cas où ces environnements sont mal jugés par leurs usagers, il reste à trouver un moyen de les rendre plus agréables, ou moins gênants. Une technique fréquemment utilisée pour cela est le masquage, dont le but est de masquer, dissimuler ou encore habiller l'environnement sonore (ou le bruit induit par une seule source sonore) afin de le rendre plus supportable ou appréciable. Le terme "masquage" est souvent utilisé de manière très globale, mais il est possible d'identifier plusieurs approches de masquage selon l'objectif posé. Par exemple, si le but de l'opération de masquage est de couvrir un son, il est plus que pertinent de faire référence au livre \textit{"Psychoacoustics : Facts and models"} de Zwicker et Fastl \cite{fastl_psychoacoustics_2007}, qui présente les propriétés d'un masquage dit "énergétique" visant à couvrir un son pur avec plusieurs types de "masques" (bruit large-bande, son pur, ou encore son complexe). \\

Dans un contexte plus appliqué, le masquage peut être vu comme solution pour traiter les problématiques de bruit. C'est ce que tentent de faire Cai et coll. dans leur étude sur la gêne induite par le bruit industriel \cite{cai_effect_2019}, dans laquelle aucune approche de masquage spécifique n'est explicitée, mais dont l'objectif est de tester l'efficacité de l'ajout de sons d'eau sur la gêne perçue. La volonté n'est donc pas forcément de couvrir le bruit, mais plutôt de l'"habiller", les sons d'eau s'apparentant plus à des "maquilleurs" qu'à des "masques". Il a ainsi été montré que la réduction de la gêne perçue dépendait du rapport de niveau entre le bruit et le maquilleur, et que cette réduction était maximale lorsque le son présenté aux participants était composé du bruit de soudeuse électrique et du son de fontaine superposés, tous les deux au même niveau. L'approche d'"habillage", ici mise en œuvre par combinaison de la source de bruit et des "maquilleurs", est d'autant plus intéressante qu'elle peut permettre de ne pas ajouter d'objet sonore à la scène considérée. En effet, en créant ou choisissant un "maquilleur" qui se fond avec le bruit, il serait possible de garder un environnement sonore homogène, qui ne capte pas l'attention de l'auditeur. \\

Au contraire, capter l'attention peut tout aussi bien constituer l'objectif d'une approche de masquage, alors appelée masquage "informationnel". Il s'agit alors de faire diversion avec le signal ajouté, pour que l'attention de l'auditeur se focalise sur autre chose que sur la source de bruit. Dans le cas de l'étude de Cai et coll. \cite{cai_effect_2019}, la combinaison du bruit de soudeuse électrique et des sons d'eau peut faire en sorte que l'attention des auditeurs soient captée par les sons d'eau, et non plus par le bruit. Cette méthode peut en revanche s'avérer non pertinente dans certaines situations, comme celle de travailleurs qui se plaignent d'un bruit qui les empêchent de se concentrer. Ajouter un autre son qui capte leur attention pourrait en effet amoindrir le problème de gêne, mais ne serait pas forcément favorable à la concentration. \\

Plusieurs approches de masquage comme celles décrites ici peuvent finalement se trouver couplées à l'évaluation du confort, de la gêne ou de l'agrément, et constituer de bonnes pistes pour traiter la problématique du bruit au niveau perceptif. 

\subsection{\textit{Best-Worst Scaling} : une méthode pour évaluer la perception sonore}\label{BWS}

Pour mener des expériences dont l'objectif est d'évaluer la perception sonore, plusieurs méthodes ont été développées. L'une d'entre elles, nommée \textit{Rating Scale (RS)}, est largement utilisée et propose une échelle sémantique permettant aux participants de juger les sons qui leur sont proposés selon un critère fixé. Si cette méthode permet de recueillir un certain nombre d'informations (le jugement des participants, mais aussi les valeurs moyennes obtenues pour un seul son ou encore la cohérence inter et intra-participants), elle présente certains désavantages. Pour que les participants se fassent une bonne représentation du corpus de sons et sachent bien les placer sur l'échelle sémantique, il est nécessaire de leur présenter les limites de ce corpus par rapport au critère étudié. Par exemple, si l'expérience a pour but d'étudier la perception de la rugosité, le fait de présenter aux participants le son le moins rugueux et le son le plus rugueux avant de commencer l'expérience permet d'assurer une bonne utilisation de l'échelle sémantique. Cependant, cela suggère que le critère étudié est défini de manière objective. Si l'évaluation concerne un critère subjectif, comme l'agrément ou la gêne perçue, il n'est pas possible pour l'expérimentateur de définir à l'avance deux sons représentant les limites du corpus, à moins de se baser sur sa propre perception. L'utilisation de l'échelle sémantique devient alors difficile à envisager, et le développement d'autres méthodes d'évaluation pour ce genre de critère semble nécessaire. Parmi les nouvelles méthodes de plus en plus utilisées, la méthode \textit{Best-Worst Scaling (BWS)} représente une bonne alternative à la méthode \textit{RS} pour l'évaluation des sons. Dans un contexte général (c'est-à-dire pas seulement sonore), cette méthode consiste à présenter plusieurs groupes d'éléments, ce qui peut se révéler très utile dans plusieurs domaines de recherche pour évaluer un grand nombre d'éléments. 

\subsubsection{Fonctionnement de la méthode \textit{Best-Worst Scaling}}
Lors de son travail visant à tester de nouveaux algorithmes capables de noter des centaines voire des milliers d'éléments, Geoff Hollis présente le fonctionnement de la méthode \textit{BWS} \cite{hollis_scoring_2018} comme suit : pour un total de $S$ éléments, les participants se voient présenter plusieurs groupes de $N$ éléments. Pour chaque groupe, ceux-ci doivent juger tous les éléments et en choisir deux qu'ils considèrent comme le "meilleur" et le "pire" selon un critère fixé. Cette technique permet d'associer un rang à chaque élément (même ceux qui ne sont jamais jugés comme meilleur ou pire) à l'aide de différents algorithmes de notation, et de générer un classement des éléments selon le critère étudié. Par exemple, en prenant $N = 4$ (ce qui est souvent le cas en pratique) et en notant $A$, $B$, $C$ et $D$ les éléments d'un groupe, il est possible de tirer cinq informations du jugement effectué par un participant pour ce groupe. Si celui-ci choisit $A$ comme étant le meilleur élément et $D$ comme le pire, il est possible de déduire les relations suivantes entre les éléments (l'inégalité $A > B$ étant interprétée comme "A est meilleur que B") : 
$$
\begin{cases}
    A > B \\
    A > C \\
    A > D \\
    B > D \\
    C > D 
\end{cases}
$$

Seule la relation entre $B$ et $C$ reste inconnue. \\

En comparaison avec la méthode \textit{RS}, la méthode \textit{BWS} permet d'évaluer un grand nombre d'éléments en seulement quelques essais (un essai correspondant à un jugement sur un groupe de $N$ éléments). Le fait d'évaluer les éléments en groupe assure également le caractère relatif du jugement, ce qui peut s'avérer compliqué avec la méthode \textit{RS} si les éléments sont présentés un par un. Néanmoins, certaines contraintes sont à prendre en compte, notamment le fait que le nombre total d'éléments $S$ doit être multiple de $N$ (ou $N$ diviseur de $S$) pour la constitution de groupes contenant le même nombre d'éléments. De plus, il n'est pas possible de faire de moyenne sur un élément avec la méthode \textit{BWS}, car les données récoltées sont des valeurs non-numériques (l'élément jugé comme meilleur et celui jugé comme pire). Il est alors possible de faire appel à des méthodes de comptage, pour calculer par exemple la fréquence à laquelle chaque élément a été choisi comme le meilleur. Il reste néanmoins possible d'utiliser la méthode \textit{BWS} dans beaucoup de domaines différents, et notamment pour le son. C'est d'ailleurs pour tester la validité d'une telle méthode sur des stimuli sonores que Rosi et coll. ont mené une expérience avec vingt participants, à qui il a été demandé d'évaluer un ensemble de $S = 100$ sons instrumentaux en termes de brillance \cite{rosi_best-worst_2022}. Les deux méthodes \textit{BWS} et \textit{RS} ont été utilisées, et les résultats ont montré que la méthode \textit{BWS} est équivalente à la méthode \textit{RS} en termes de performance, et plus rapide que cette dernière. De plus, les participants ont préféré la méthode \textit{BWS}, même s'il est selon eux plus difficile de faire un choix dans le cas où les sons présentés sont très similaires. La méthode \textit{BWS} semble néanmoins un paradigme adapté pour l'étude de la perception sonore, et pourrait présenter une alternative à la méthode \textit{RS} pour l'évaluation de critères subjectifs comme l'agrément.

\subsubsection{Algorithmes de notation : mise à jour des scores d'un grand nombre d'éléments}
Lors d'une expérience faisant appel à la méthode \textit{BWS}, un certain nombre d'éléments sont comparés et jugés les uns par rapport aux autres selon un critère énoncé. Une fois les jugements des participants recueillis, plusieurs algorithmes peuvent être utilisés pour calculer le rang (ou le score) de chaque élément. Certains d'entre eux s'étant révélés coûteux en termes de temps de calcul, Hollis met en avant la nécessité de développer de nouveaux algorithmes qui puissent convertir les jugements de la méthode \textit{BWS} en scores pour un grand nombre d'éléments. Il présente alors dans son étude le fonctionnement de différents algorithmes de notation, dont les performances sont validées empiriquement grâce à des simulations. Parmi ces algorithmes, trois ont révélé de bons résultats lors des différentes expériences \cite{hollis_scoring_2018}. \\

Le premier est l'algorithme de notation des tournois développé par Arpad Elo \cite{elo1973international}, dont le fonctionnement s'apparente à un tournoi entre les éléments, qui sont vus comme autant de concurrents qui s'affrontent lors de matchs. Lorsqu'un participant est confronté à un groupe de $N$ éléments et qu'il doit en choisir un meilleur et un pire, il est alors possible d'imaginer un tournoi entre les $N$ éléments. Les relations entre les éléments représentées par les inégalités ci-dessus correspondent alors aux issues de certains matchs, et l'inégalité $A>B$ peut être interprétée comme "victoire du concurrent $A$ face au concurrent $B$", ce qui correspond à "l'élément $A$ a été jugé meilleur que l'élément $B$". Pour chacune de ces relations, le score des deux concurrents concernés est mis à jour en fonction de l'écart entre leurs scores avant le match : un concurrent avec un score élevé voit son score augmenter légèrement s'il gagne contre un concurrent avec un score bas, tandis qu'un concurrent avec un score bas voit son score augmenter beaucoup s'il gagne contre un concurrent avec un score élevé. Cette méthode prend donc en compte non seulement le résultat d'un jugement (ou d'un match), mais aussi l’écart de "compétence" entre les concurrents, c'est-à-dire la probabilité de chaque concurrent de gagner face à un autre. Ainsi, pour un concurrent $A$, son rang $R_A$ est mis à jour après un match en fonction de l'écart entre son score réel $S_A$ et son score attendu $E_A$, multiplié par un facteur $K$, comme présenté dans l'équation \eqref{eq1}.
\begin{equation}
R'_A = R_A + K(S_A - E_A)
\label{eq1}
\end{equation}

Pour déterminer le score attendu d'un concurrent, le système de notation Elo suppose que la compétence des concurrents suit une distribution normale avec un écart-type de 100 sur l’échelle Elo, et que c'est cette compétence qui détermine principalement l'issue des matchs entre deux concurrents. Par conséquent, pour deux concurrents $A$ et $B$ de compétences égales, il est supposé que $A$ et $B$ gagnent l’un contre l’autre un nombre égal de fois. À l'inverse, si $A$ possède un score plus élevé que $B$, alors $A$ est plus susceptible de gagner. À partir de ces hypothèses, la probabilité attendue qu’un concurrent $A$ gagne contre un concurrent $B$ est définie par l'équation \eqref{eq2} \cite{elo1973international} : 
\begin{equation}
E_A = \frac{10^{\frac{R_A}{400}}}{10^{\frac{R_A}{400}} + 10^{\frac{R_B}{400}}}
\label{eq2}
\end{equation}

La règle de mise à jour de l'algorithme Elo est sensible à la différence de rang entre les deux éléments concernés. De plus, la complexité pour les mises à jour des scores est constante, ce qui ne limite pas cet algorithme en termes d’efficacité computationnelle, et le rend pertinent lorsque de nombreux éléments doivent être notés. \\

Un deuxième algorithme, cette fois dit "discriminant", est basé sur le modèle de Rescorla-Wagner \cite{rescorla1972theory}, qui met à jour des associations entre des indices et des événements. Lorsqu’un indice survient (par exemple un concurrent particulier) et est suivi d’un événement spécifique (par exemple ce concurrent gagne un match face à un autre), la force d’association $V_x^{n}$ entre l’indice et l’événement est mise à jour pour refléter la relation observée grâce à l'équation \eqref{eq3} suivante : 
\begin{equation}
V_x^{n+1} = V_x^{n} + \Delta V_x^{n+1}
\label{eq3}
\end{equation}

Le degré de changement dû à chaque paire indice-événement $\Delta V_x^{n+1}$ dépend de plusieurs paramètres : 

- la saillance de l’association observée notée $\alpha$,

- le taux d’apprentissage pour l’indice noté $\beta$, 

- la force d’association maximale pour l’événement notée $\lambda$,

- la force d’association totale entre l’indice et tous les événements, notée $V_{tot}$. \\

Les équations \eqref{eq4} et \eqref{eq5} suivantes permettent respectivement d'exprimer ce degré de changement, ainsi que la saillance d'une association : 
\begin{equation}
\Delta V_x^{n+1} = \alpha_x \beta (\lambda - V_{tot})
\label{eq4}
\end{equation}

\begin{equation}
\alpha = 1 - \frac{V_w(meilleur)}{V_w(meilleur) + V_w(pire)} 
\label{eq5}
\end{equation}

La règle de mise à jour de Rescorla-Wagner peut intégrer des informations comme celle de l'algorithme Elo sur la qualité d'un jugement via le terme de saillance (lorsqu’un élément $A$ gagne contre un élément $B$ avec un score beaucoup plus faible, le résultat est moins saillant que si $B$ avait un score beaucoup plus élevé). L'équation \eqref{eq5} suppose que les événements inattendus sont saillants, en estimant la probabilité que $A$ gagne contre $B$ comme $\frac{V_w(A)}{V_w(A) + V_w(B)}$, avec $V_w(A)$ la force d'association entre $A$ et l'événement "$A$ gagne le match", et $V_w(B)$ la force d'association entre $B$ et l'événement "$A$ gagne le match". Dans le cas où aucun des éléments n’a d’association préalable avec l'événement cité, le terme de saillance est fixé à $0.5$. En comparaison avec l'algorithme de notation Elo, qui donne des scores qui n'ont pas de limites supérieures ou inférieures (tant qu'un élément particulier gagne de manière cohérente, des distinctions de plus en plus grandes peuvent être faites entre lui et le reste des concurrents), les forces d'association du modèle de Rescorla-Wagner sont limitées par une force de conditionnement maximale théorique possible entre un indice et un événement, spécifiée par le paramètre $\lambda$. \\

Enfin, un troisième algorithme de notation appelé \textit{Value Learning} est introduit, qui consiste en l’apprentissage de l'issue attendue d’un match pour chaque concurrent. Le classement du tournoi peut ensuite être déterminé en ordonnant les concurrents en fonction des valeurs de leur score attendu lorsqu'ils participent à un match. La valeur du score du concurrent $A$, notée $V_A$, est calculée en mettant à jour le score estimé du concurrent en fonction des résultats observés lors des matchs auxquels il participe (représentés par le paramètre $\gamma$), comme présenté par l'équation \eqref{eq6}. 
\begin{equation}
V_A = V_A + \alpha \beta (\gamma - V_x) 
\label{eq6}
\end{equation}

La règle de mise à jour de l'algorithme \textit{Value Learning} est similaire à celle de Rescorla-Wagner, car les scores sont déduits à partir de la différence entre ce qui était attendu et ce qui a réellement été observé. Cependant, \textit{Value Learning} se concentre sur l'issue observée d'un événement (victoire ou défaite), plutôt que sur une association entre un indice et un événement. Les mêmes paramètres sont néanmoins retrouvés, à savoir le taux d’apprentissage $\beta$ et la saillance $\alpha$. Ainsi, les chances de victoire $O_A$ du concurrent $A$ sont calculées en considérant son score attendu comme une probabilité de victoire, et en la convertissant en une valeur de cote, comme le montre l'équation \eqref{eq7}. La saillance de l'issue est ensuite calculée selon l'équation \eqref{eq8}, qui assure que celle-ci soit entre 0 et 1, et qu'elle augmente à mesure que le résultat observé devient inattendu. Dans le cas où deux concurrents n’ont encore jamais gagné de match, la saillance est fixée à $0.5$.

\begin{equation}
O_A = \frac{V_A}{1 - V_A}
\label{eq7}
\end{equation}

\begin{equation}
\alpha = 1 - \frac{O_A}{O_A+O_B}
\label{eq8}
\end{equation}

Une des limitations de cette méthode est que les valeurs des différentes issues des matchs (victoire, défaite, etc.) doivent être définies à l'avance. Toutefois, la valeur absolue de ces résultats n'est pas ce qui importe le plus; c'est la relation entre les différentes valeurs qui est essentielle : qu'une victoire vaille $1$ ou $10$ n'a pas d'importance tant que la défaite vaut $0$ et qu'un match nul se situe quelque part entre les deux. \\

Finalement, chacun de ces trois algorithmes permet d'évaluer des scores pour un grand nombre d'éléments. Dans le contexte particulier d'une étude impliquant la méthode \textit{BWS}, les simulations réalisées par Hollis pourront se révéler utiles pour le choix de l'algorithme de notation le plus efficace et le plus adapté. Une fois les scores calculés, ceux-ci peuvent être interprétés en relation avec les caractéristiques des éléments notés, et des tests statistiques peuvent être utilisés sur des groupes d'éléments pour confirmer ou réfuter certaines hypothèses de recherche.

\newpage
\section{Construction d'une nouvelle approche de masquage}\label{section3}
Les notions qui serviront de socle à l'axe \textit{Auditory Concealer} étant maintenant définies, la question est de savoir comment les mettre en œuvre dans le cadre du projet \textit{ReNAR}. Cette partie détaillera toute la réflexion qui s'est menée autour de ces notions et présentera les aspects théoriques et pratiques de l'approche \textit{concealer}, dans le but de la comparer à l'approche \textit{masker}.

\subsection{Objectifs et enjeux du projet \textit{ReNAR}}\label{objectifs}
Au vu de l'objectif premier du projet \textit{ReNAR}, qui est de proposer une solution à l'exposition au bruit des personnes dans leurs espaces de vie, la notion de confort a été identifiée comme notion centrale du projet, et guidera la réflexion autour du bruit dans les bureaux ouverts. Aussi, le projet n’ayant pas pour visée de réduire physiquement le bruit mais bien de modifier la perception des usagers, le mot “confort” sera utilisé dans la suite pour parler de “confort sonore”, tel que défini dans l'état de l'art (section \ref{etat-art}). \\

Aussi, ayant choisi d'étudier le bruit dans les environnements spécifiques que sont les bureaux ouverts, il est important de garder à l'esprit que le confort est étroitement lié à différents éléments de contexte correspondant à ce type d'environnement. Dans ce cadre d'étude, la tâche effectuée par les usagers, les sources constituant l'ambiance sonore du lieu, ou encore la sensibilité au bruit de chacun sont autant d'éléments de contexte à prendre en compte et à ne pas négliger. Dans le modèle de confort de Vink et Hallbeck \cite{vink_editorial_2012} repris par Duroyon \cite{duroyon2025confort} présenté en figure \ref{modele-confort}, une place importante est donnée aux attentes de l'utilisateur, car celles-ci influent sur son jugement du produit étudié (jugement séparé en trois cas dans le modèle : confort, inconfort, et neutre). De plus, ce modèle suggère l'étude de produits comme objets manufacturés, ce qui ne correspond pas au cadre du projet \textit{ReNAR}, qui traite plutôt d'un environnement. Dans le cas particulier des bureaux ouverts, il est néanmoins possible d'assimiler un environnement à un produit, dont les usagers sont des travailleurs avec certaines attentes. Celles-ci pourraient par exemple se formuler de la manière suivante : "je m'attends à ce que mon environnement de travail soit propice à la concentration et au calme", ou bien en ajoutant plus de contexte : "en temps que membre d'une équipe, je m'attends à ce que la communication soit facilitée sur mon lieu de travail". L'évaluation du confort concerne alors l'ambiance sonore de l'environnement étudié dans son ensemble, et ce en cohérence avec le contexte dans lequel cette ambiance est soumise aux usagers de l'espace. \\

Cela étant dit, le premier axe \textit{Auditory Concealer} du projet \textit{ReNAR} se propose de développer une nouvelle approche de masquage, nommée approche \textit{concealer}, dont le but est de modifier les environnements sonores perçus par les usagers de différents espaces en habillant les sons indésirables (c'est-à-dire jugés gênants ou désagréables). Cette approche de réalité augmentée sonore jouera pour cela sur les propriétés spectro-temporelles des sons, et devra s'efforcer de minimiser l'augmentation du niveau sonore global de la scène étudiée. En effet, le fait que le niveau sonore soit le premier facteur responsable de la gêne et du désagrément impose une attention particulière sur ce paramètre, qui constituera un facteur central lors de la construction de l'approche \textit{concealer} dans les parties suivantes.

\subsection{Définition des approches \textit{masker} et \textit{concealer}}\label{approches}
Dans le cadre d'environnements de travail, il est intéressant de s'appuyer sur l'étude menée par Cai et coll., qui vise à évaluer l'effet du masquage sur la gêne induite par un bruit industriel \cite{cai_effect_2019}. L'approche de masquage utilisée vise à masquer une source de bruit en y ajoutant certains sons, afin de créer une mixture qui se veut moins gênante, ou plus agréable. Une telle approche est en ce sens appelée approche \textit{masker}, et se résume par le schéma présenté sur la figure \ref{approche-masker} suivante : 

\begin{figure}[h!]
    \centering
    \includegraphics[scale = 0.55]{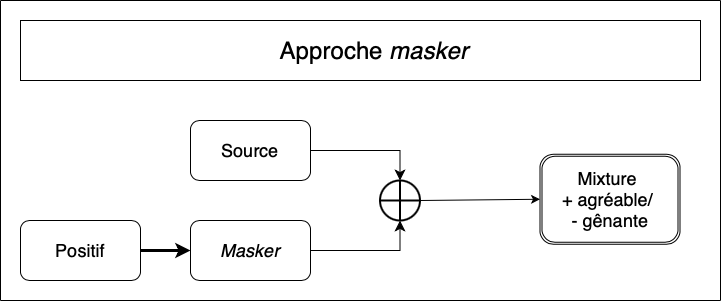}
    \caption{Schéma résumant le principe de l'approche masker}
    \label{approche-masker}
\end{figure}

La scène sonore résultante de l'application de cette approche sur une source de bruit est alors une mixture entre ce bruit et un signal ajouté nommé \textit{masker}. Ce dernier est construit à partir d'un son qualifié de "positif", c'est-à-dire considéré comme communément accepté voire apprécié. Dans le cas de l'étude de Cai et coll., la source choisie est un bruit de soudeuse électrique, et des sons d'eau (fontaine, pluie et cascade) sont utilisés comme \textit{maskers}, car ceux-ci sont a priori considérés comme des sons positifs. L'addition se fait dans le domaine temporel, et il est intéressant de jouer sur le niveau sonore du \textit{masker} pour trouver un compromis entre sa présence perçue dans la mixture et le niveau sonore global de la scène. Cai et coll. utilisent pour cela un rapport de niveau entre le bruit et le \textit{masker}, appelé \textit{Masking Sound to Noise Ratio (MSNR)}. Dans le cas de l'approche \textit{masker}, le \textit{"Masking Sound"} est le signal \textit{masker}, et le \textit{"Noise"} la source de bruit. De plus, le \textit{MSNR} est exprimé en dB(A), de sorte qu'il soit positif lorsque le bruit est moins fort que le \textit{masker}. \\

Pour l'approche \textit{concealer}, le raisonnement est différent : l'idée est plutôt de compléter la source de bruit. Le signal ajouté, cette fois appelé \textit{concealer}, est construit à partir du même positif que pour l'approche \textit{masker}, mais aussi à partir de la source. Source et \textit{concealer} sont ensuite ajoutés pour donner une mixture qui se veut elle aussi plus agréable et moins gênante. Cette nouvelle approche est présentée sur la figure \ref{approche-concealer} ci-dessous : 

\begin{figure}[h!]
    \centering
    \includegraphics[scale = 0.55]{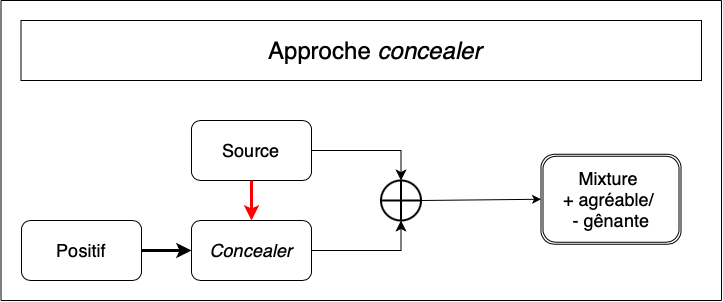}
    \caption{Schéma résumant le principe de l'approche concealer}
    \label{approche-concealer}
\end{figure}
\newpage

Contrairement à l'approche \textit{masker}, l'approche \textit{concealer} permettrait de compléter une source pour garantir une mixture plus acceptée, voire agréable si le positif est bien choisi, mais aussi de travailler à des niveaux ajoutés moins importants. Puisque le signal \textit{concealer} dépend de la source, il est également important de prendre en compte le contenu fréquentiel de celle-ci ainsi que celui du positif pour construire un bon \textit{concealer}. Par exemple, si la source à habiller est le bruit d'un système de ventilation, il est difficilement concevable de choisir un chant d'oiseau composé uniquement de hautes fréquences comme positif, car le \textit{concealer} se trouverait lui aussi composé de hautes fréquences, rendant la mixture entre lui et la source très hétérogène. En revanche, un son de vagues, dont le contenu fréquentiel est similaire à celui du son de ventilation, peut représenter une bonne option pour le positif, rendant une mixture plus homogène. À la création du \textit{concealer}, les sons de ventilation et de vagues peuvent en effet se compléter l'un l'autre. La question du choix des positifs est donc primordiale pour que l'approche \textit{concealer} permette d'améliorer le confort dans certains environnements, dont celui des bureaux ouverts. \\

Finalement, concernant les deux approches \textit{masker} et \textit{concealer}, le point essentiel est de construire des mixtures mieux perçues que le bruit. Cela requiert une bonne définition des signaux \textit{masker} et \textit{concealer}, tous les deux dépendant du positif. Une première étape consiste alors à s'assurer que le positif est effectivement perçu comme agréable, et ce dans un contexte bien défini. La perception d'un son peut en effet changer selon l'environnement et la situation dans lesquels il sont expérimentés. Une musique avec paroles peut par exemple paraître agréable dans un contexte de loisirs, mais être perçue comme moins agréable voire gênante dans un contexte de travail.

\subsection{Choix de la source et du positif}\label{choix-sons}
Maintenant que les deux approches sont théoriquement bien définies, il faut être en mesure de les implémenter. Pour cela, il est nécessaire de connaître la source de bruit et le positif, qui permettront de créer les signaux \textit{masker} et \textit{concealer} (ceux-ci pourront être regroupés sous le terme de "maquilleur" pour parler du signal ajouté indépendamment des deux approches). Ce faisant, les différents signaux seront notés de la manière suivante dans la suite : 

- $s$, la source de bruit;

- $m$, le \textit{masker};

- $c$, le \textit{concealer};

- $p$, le positif.

\subsubsection{Choix de la source}\label{source}
Pour ce qui est de la source, il a été décidé d'étudier le bruit d'un système de ventilation. Dans le contexte des bureaux ouverts, cette source sonore est en effet très pertinente par sa présence quasiment permanente dans ce type d'espace. De plus, le caractère stationnaire de ce bruit a permis de réaliser de premiers essais assez simples avec les deux approches de masquage (un bruit présentant de grandes variations d'amplitude ou dans son contenu fréquentiel aurait en effet été beaucoup plus difficile à traiter dans un premier temps). \\

\begin{figure}[h!]
    \centering
    \includegraphics[scale = 0.4]{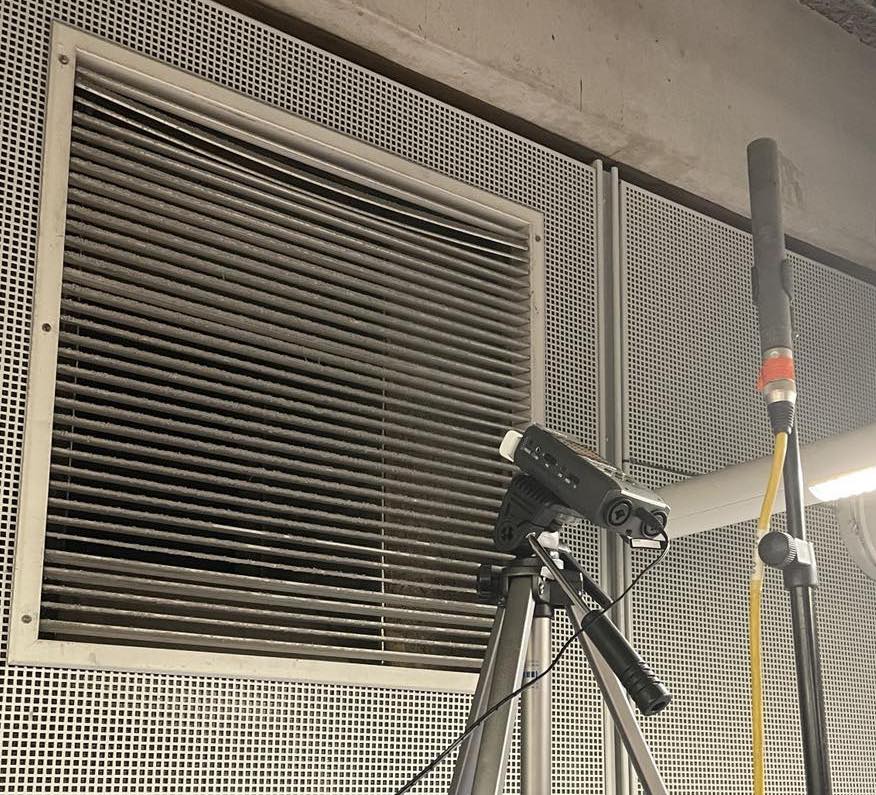}
    \caption{Prise de son en champ proche du système de ventilation présent dans les bureaux du laboratoire STMS (équipe Perception et Design Sonores)}
    \label{rec-prox}
\end{figure}

Deux bruits de ventilation ont été sélectionnés. Le premier est le bruit du système de ventilation présent dans les bureaux du laboratoire STMS (équipe Perception et Design Sonores) au sein de l'IRCAM. Celui-ci a été enregistré de différentes manières : 

- en champ proche avec un \textit{Zoom H4N} (Figure \ref{rec-prox});

- en champ proche avec un microphone \textit{Schoeps} muni d'une capsule \textit{MK5} réglée en position cardioïde et d'un pré-ampli \textit{CMC5} (Figure \ref{rec-prox});

\begin{figure}[!h]
         \centering
         \begin{subfigure}[]{0.50\textwidth}
             \centering
             \includegraphics[scale = 0.3]{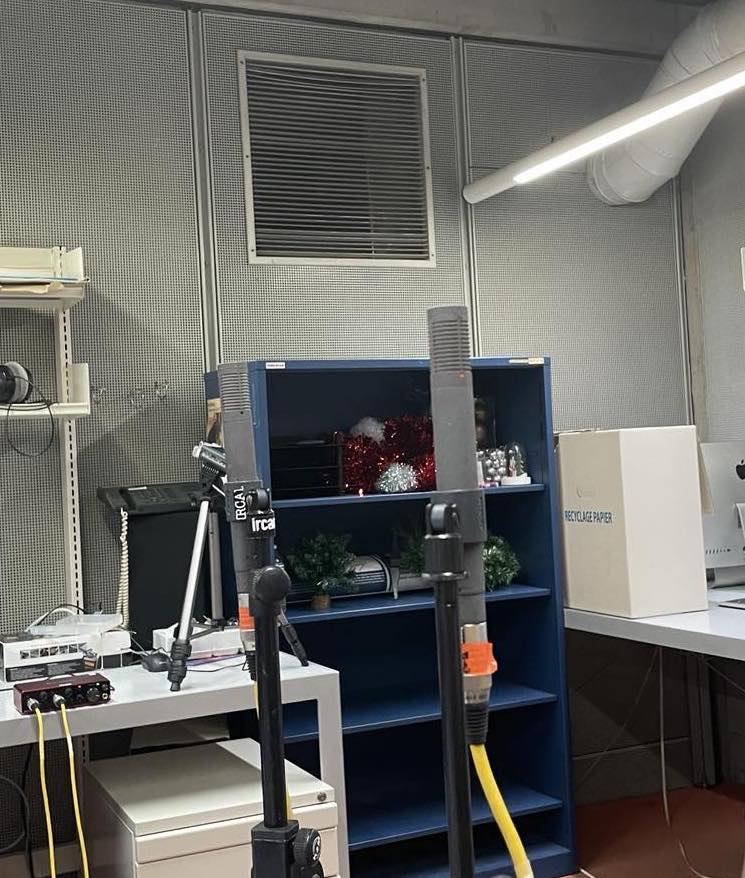}
             \caption{Vue de profil}
         \end{subfigure}
         \hspace{50pt}
         \begin{subfigure}[]{0.50\textwidth}
             \centering
             \includegraphics[scale = 0.26]{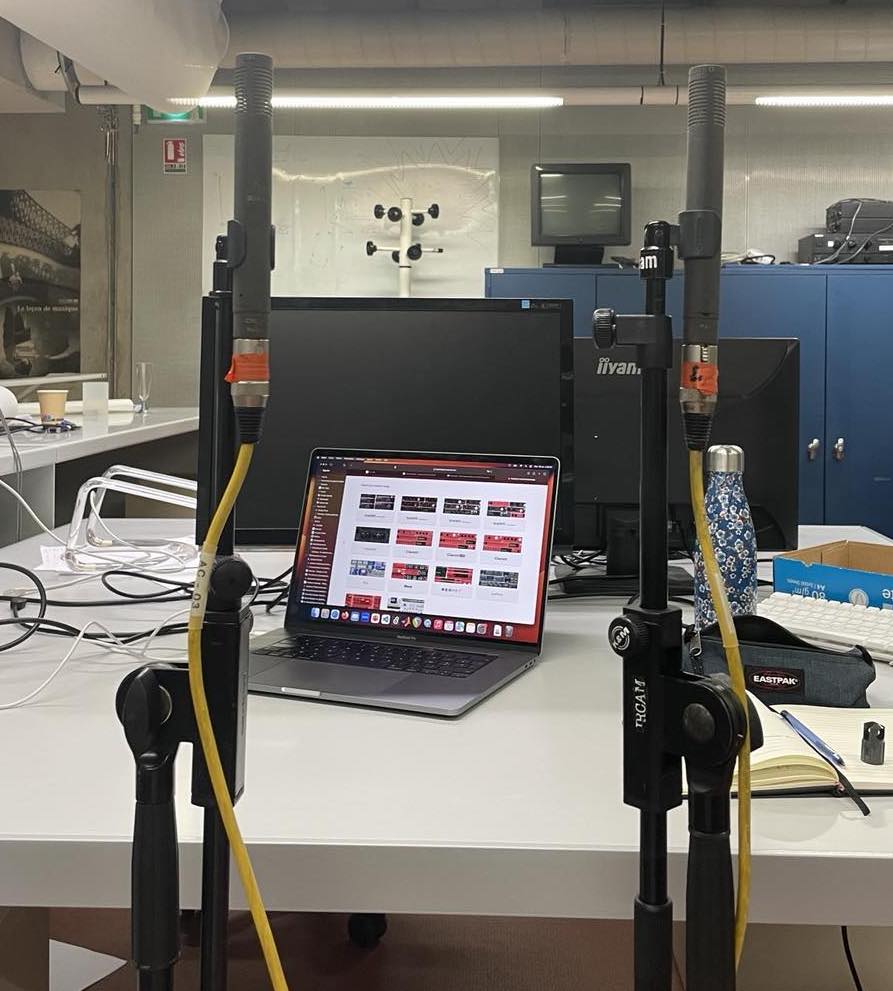}
             \caption{Vue de face}
         \end{subfigure}
        \caption{Prise de son du système de ventilation présent dans les bureaux du laboratoire STMS (équipe Perception et Design Sonores) au niveau d'un auditeur assis à son bureau}
        \label{rec-far}
\end{figure}

- à la position d'un auditeur avec deux microphones \textit{Schoeps} présentant les mêmes caractéristiques décrites précédemment (ceux-ci étant placés au niveau des oreilles de l'auditeur si celui-ci était assis à un bureau) (Figure \ref{rec-far}). 

\newpage

Une mesure au sonomètre a également été réalisée afin de caractériser le niveau auquel peuvent être soumis les usagers du bureau. Le sonomètre a pour cela été placé à la place d'un usager, et un niveau $L_{Aeq}$ moyen de $33dB(A)$ a été recueilli sur deux minutes de mesure. \\

\begin{figure}[!h]
         \centering
         \begin{subfigure}[]{0.40\textwidth}
             \centering
             \includegraphics[scale = 0.5]{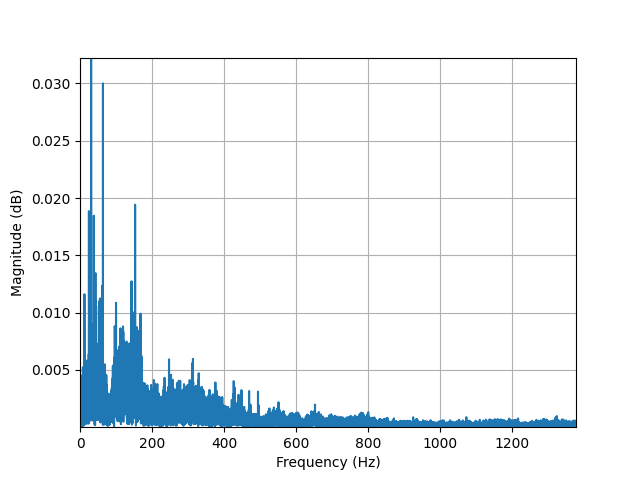}
             \caption{Spectre de "ventil1"}
         \end{subfigure}
         \hspace{50pt}
         \begin{subfigure}[]{0.40\textwidth}
             \centering
             \includegraphics[scale = 0.5]{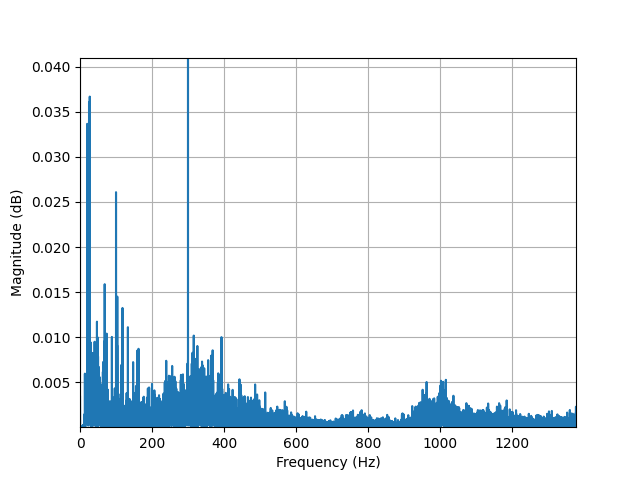}
             \caption{Spectre de "ventil2"}
         \end{subfigure}
        \caption{Spectre des deux bruits de ventilation}
        \label{sources-spectra}
\end{figure}

Par ailleurs, il a semblé que cette source en particulier n'était pas très représentative de l'aspect gênant que peuvent avoir certains bruits de ventilation. Il a donc été décidé d'en sélectionner une autre dans le corpus sonore utilisé dans l'étude de Susini et coll., qui vise à caractériser la qualité sonore des bruits de ventilation \cite{susini_characterizing_2004}. Ce deuxième bruit de ventilation présente une composante tonale continue en plus de la soufflerie basse fréquence, comme présenté sur la figure \ref{sources-spectra} \footnote{Les deux sons de ventilation sont contenus dans les fichiers \textit{10s\_ventil-ircam\_65dB(A).wav} et \textit{10s\_c12dL\_65dB(A).wav}, disponibles dans le dossier \textit{"sounds/ventilation\_noises/"}.}. \\

Le terme "source" évoqué dans les deux approches de masquage présentés précédemment (Figures \ref{approche-masker} et \ref{approche-concealer}) sera maintenant utilisé pour faire référence aux deux bruits de ventilation ainsi choisis. Pour les différencier, ceux-ci seront nommés "ventil1" pour le bruit enregistré dans le bureau de l'équipe Perception et Design Sonores, et "ventil2" pour le bruit provenant du corpus de Susini et coll. \cite{susini_characterizing_2004}.

\newpage

\subsubsection{Choix du positif}\label{positif}
Les sources ainsi définies, il a fallu déterminer le(s) positif(s) à utiliser pour la création des signaux \textit{masker} et \textit{concealer}. Pour cela, il a été décidé de s'inspirer de l'étude sur l'effet de masquage par les sons d'eau de Cai et coll. \cite{cai_effect_2019}, qui fournit des résultats pour l'approche \textit{masker}. Celle-ci a pu aiguiller la réflexion autour de la construction des mixtures avec les deux approches. Pour les sons de fontaine, de pluie et de cascade, il a en effet été montré que la mixture avec le bruit de soudeuse électrique était la moins gênante pour des valeurs de \textit{MSNR} entre $-3$ et $0$dB(A), ce qui correspond à des niveaux de sons d'eau inférieurs à celui de la source de bruit \cite{cai_effect_2019}. Le fait que les sons d'eau n'aient pas besoin d'être à de hauts niveaux pour rendre la mixture moins gênante correspond bien à l'idée de minimiser le niveau de la mixture produite par l'approche \textit{concealer}.

Par ailleurs, certains sons d'eau ont été classés comme communément acceptés par la littérature, et notamment par Maria Rådsten-Ekman lors de sa thèse et de plusieurs études visant à déterminer si les sons d'eau peuvent être utilisés pour améliorer la perception du bruit environnemental en milieu urbain \cite{radsten-ekman_may_2010} \cite{radsten-ekman_similarity_2015}. Il semble alors pertinent de les utiliser comme positifs, afin d'espérer créer des mixtures agréables, et de comparer les deux approches. \\

Cela étant dit, cinq sons d'eau ont été retenus après avoir parcouru les banques de sons disponibles à l'IRCAM. Parmi eux, des sons de pluie, de cours d'eau et de cascade (nommés \textit{rain}, \textit{stream} et \textit{waterfall} dans la suite) ont été sélectionnés dans la banque \textit{Sound Ideas} \cite{soundideas}, et des sons de fontaine et de vagues (\textit{fountain} et \textit{waves}) ont été sélectionnés dans la banque \textit{Blue Box} \cite{bluebox} \footnote{Les sons d'eau sont également disponibles dans le dossier \textit{"sounds/water\_sounds/"}, dans les fichiers \textit{10s\_fountain.wav}, \textit{10s\_rain.wav}, \textit{10s\_stream.wav}, \textit{10s\_waterfall.wav} et \textit{10s\_waves2.wav}.}. Il est également important de noter que le choix des sons d'eau a été fait dans l'optique d'une première expérience assez exploratoire. Une impasse a donc été faite sur le problème de congruence lié au contexte des bureaux ouverts. Comme précisé dans la section \ref{objectifs}, il est essentiel de choisir des sons cohérents avec l'espace dont on veut améliorer le confort. Si les sons d'eau n'entrent pas forcément dans les attentes des usagers des bureaux ouverts, il reste tout de même intéressant de les étudier pour confirmer ou non leur statut de sons positifs dans ce contexte. \\

Le choix des sons d'eau a finalement été motivé par leur diversité au niveau spectral. En plus de la comparaison globale entre les deux approches, les différents types de sons d'eau pourront en effet présenter un facteur secondaire influençant la perception des mixtures créées avec les deux approches. En particulier, cette diversité pourra fournir des informations plus précises quant aux conditions d'utilisation de l'approche \textit{concealer}.

\subsection{Construction du signal \textit{concealer}}
Les signaux \textit{maskers} correspondant simplement aux positifs choisis ($m = p$), il est d'ores et déjà possible de mettre en œuvre l'approche \textit{masker} en effectuant l'addition $s + m$ dans le domaine temporel. Mais concernant l'approche \textit{concealer}, il reste à construire $c$, qui dépend à la fois du positif et de la source. Pour rendre la mixture résultante de l'addition $s+c$ agréable, l'idée est que celle-ci s'approche le plus possible du positif. Pour cela, plusieurs méthodes ont été implémentées, impliquant chacune une construction différente du contenu fréquentiel du \textit{concealer} $c$.

\subsubsection{Définition des méthodes de construction}\label{def-methodes}
L'idée est de trouver $c$ tel que
$$s + c = \Tilde{p}$$
avec $\Tilde{p}$ une estimation du positif $p$. \\

Pour obtenir $\Tilde{p}$, il est possible d'utiliser les signaux $s$ et $p$ pour déduire simplement $c$ en opérant la soustraction 
$$c = p - s$$
En revanche, il ne faut pas oublier que ces signaux sont définis dans le domaine temporel. Or, pour tenter d'approcher $p$, il est nécessaire de jouer sur la structure spectro-temporelle de $c$, afin d'ajouter le contenu fréquentiel nécessaire à cette reconstruction. Les signaux $s$ et $p$ sont donc placés dans le domaine temps-fréquence ($s \rightarrow S$ et $p \rightarrow P$) pour déterminer l'amplitude et la phase du \textit{concealer} $C$. Il suffira ensuite de repasser $C$ dans le domaine temporel pour obtenir $c$ et l'ajouter à la source. \\

Pour calculer l'amplitude $m_C$ et la phase $a_C$ du \textit{concealer}, une transformée de Fourier à court-terme (TFCT) est calculée pour les signaux $s$ et $p$ grâce à la bibliothèque \textit{librosa} disponible sous \textit{Python} \cite{mcfee2015librosa}. L'amplitude du \textit{concealer} correspond alors à la soustraction entre les amplitudes du positif et de la source, et sa phase est fixée à la phase du positif, comme suit :
$$
\begin{cases}
    m_C = m_P - m_S\\
    a_C = a_P 
\end{cases}
$$
avec $m$ et $a$ désignant respectivement l'amplitude (magnitude) et la phase (angle) des signaux d'une part, et $S$, $P$ et $C$ désignant respectivement les représentations spectrales des signaux source, positif, et \textit{concealer} d'autre part. \\

Au vu des équations ci-dessus, un problème se présente au niveau des amplitudes : pour certaines fréquences, il est possible d'avoir $m_P > m_S$, et donc $m_C < 0$, ce qui est mathématiquement incorrect puisque l'amplitude est définie comme la valeur absolue du spectre d'un signal. Plusieurs méthodes ont été envisagées pour pallier ce problème, et construire une amplitude $m_C$ positive. \\

Avant de chercher à satisfaire la condition $m_C \geq 0$, une première méthode a été définie de sorte à construire le signal \textit{concealer} avec une amplitude qui peut être négative. Si cela peut paraître absurde d'un point de vue mathématique, il a tout de même été intéressant d'écouter le résultat après retour dans le domaine temporel. Un exemple de la forme de l'amplitude d'un tel \textit{concealer} est représentée en vert sur la figure \ref{method-non} ci-dessous : 

\begin{figure}[h!]
    \centering
    \includegraphics[scale = 0.5]{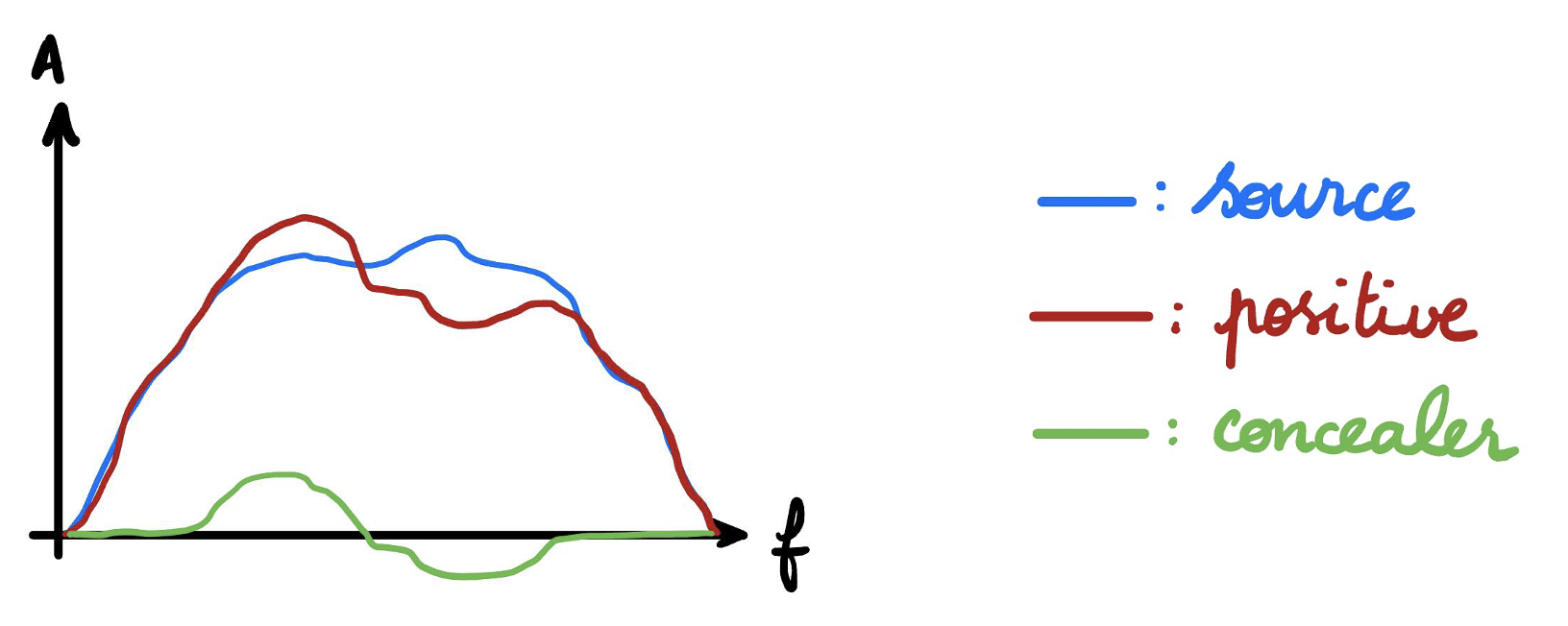}
    \caption{Schéma présentant l'amplitude du signal concealer en fonction de la fréquence (en vert), résultat de la soustraction des amplitudes du positif (en rouge) et de la source (en bleu)}
    \label{method-non}
\end{figure}

Cette première méthode de construction a permis de réaliser de premiers essais et de déterminer $m_C$ pour les sources et les positifs choisis en sections \ref{source} et \ref{positif}. \\

Afin d'éviter une amplitude négative, une deuxième méthode a cette fois consisté à appliquer une valeur absolue à la soustraction, de sorte que $m_C = |m_P - m_S|$, comme présenté sur la figure \ref{method-abs} suivante :

\begin{figure}[h!]
    \centering
    \includegraphics[scale = 0.5]{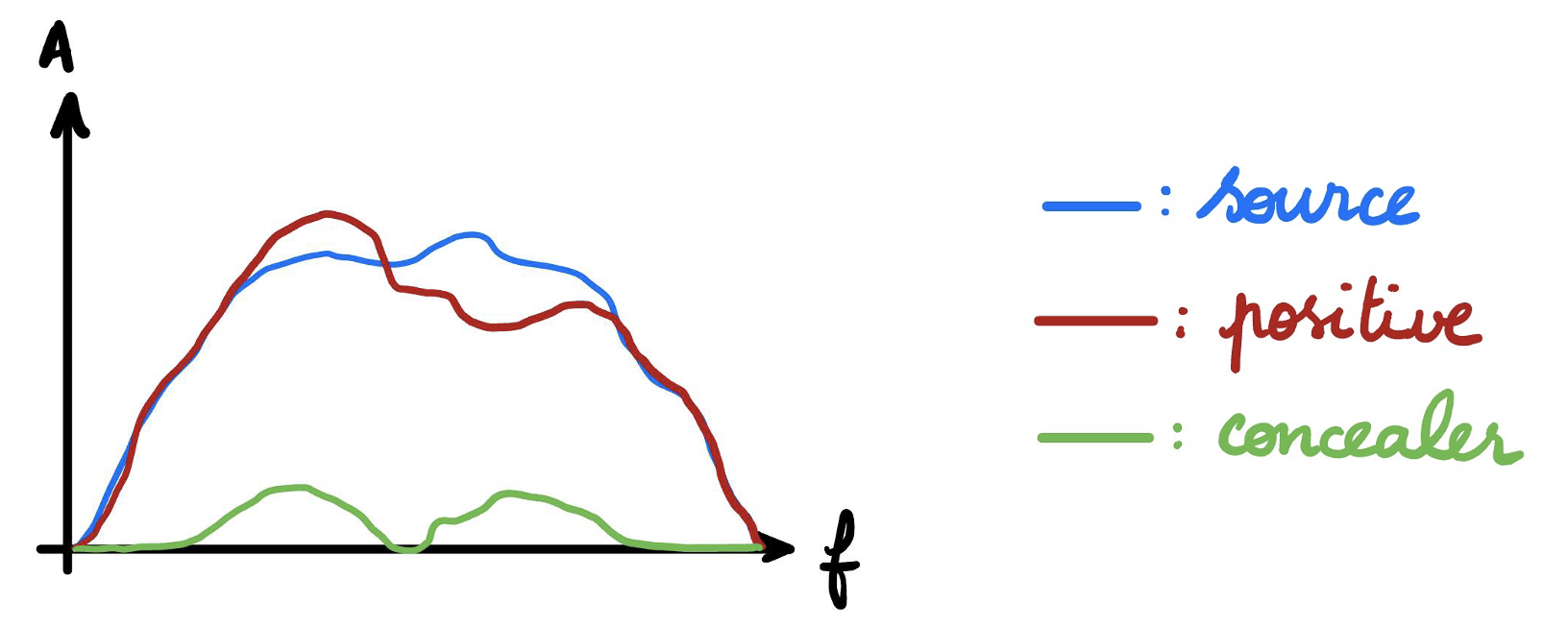}
    \caption{Schéma présentant l'amplitude du signal concealer en fonction de la fréquence (en vert), résultat de la valeur absolue de la soustraction des amplitudes du positif (en rouge) et de la source (en bleu)}
    \label{method-abs}
\end{figure}

Cette construction a permis de garantir une amplitude positive, mais ne représente pas réellement les dynamiques entre l'amplitude du positif et celle de la source. En effet, dans le cas où $m_P < m_S$, il y a moins d'énergie dans le spectre du positif que dans celui de la source. Il n'est donc pas possible d'approcher le spectre du positif en ajoutant du contenu fréquentiel à la source, comme le suggère la figure \ref{method-abs} ci-dessus (le fait d'avoir $m_C > 0$ signifiant un ajout de contenu fréquentiel à la source). \\

\begin{figure}[h!]
    \centering
    \includegraphics[scale = 0.5]{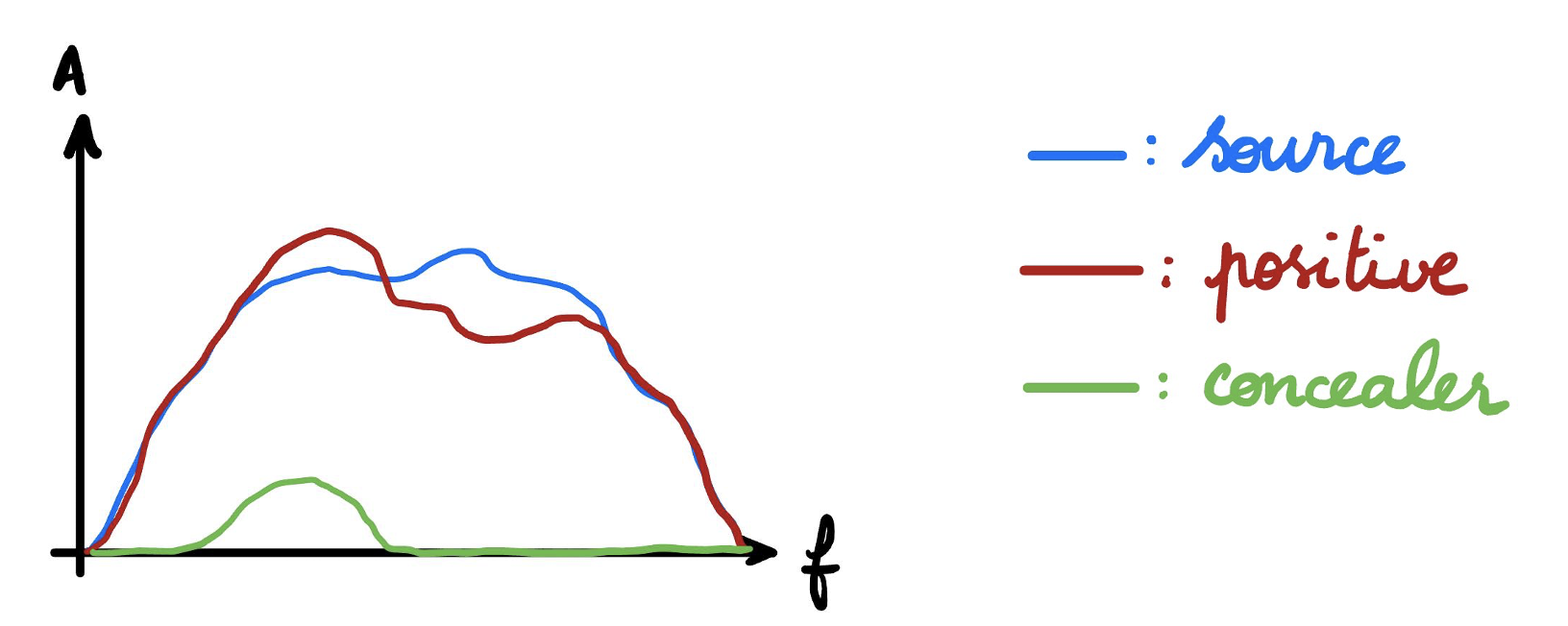}
    \caption{Schéma présentant l'amplitude du signal concealer en fonction de la fréquence (en vert), résultat de la soustraction non négative des amplitudes du positif (en rouge) et de la source (en bleu)}
    \label{method-relu}
\end{figure}

Pour rectifier cela, une troisième et dernière méthode de construction du \textit{concealer} a été envisagée, qui consiste à appliquer la soustraction $m_P - m_S$ seulement lorsque $m_P \geq m_S$. L'amplitude $m_C$ est alors définie comme suit : 
$$
m_C =
\begin{cases}
    m_P - m_S & \mbox{si } m_P \geq m_S \\
    0 & \mbox{sinon.}
\end{cases}
$$
Avec une telle méthode, une amplitude positive est garantie, et aucun ajout de contenu fréquentiel n'est effectué lorsque $m_P < m_S$, comme présenté sur la figure \ref{method-relu} ci-après. 

Un inconvénient à cette méthode reste le fait qu'en pratique, le spectre de la source et du positif n'est jamais aussi lisse que sur la figure \ref{method-relu}. Si la différence $m_P - m_S$ change très souvent de signe, le spectre du signal \textit{concealer} contiendra beaucoup de zéros. \\

Finalement, grâce aux trois méthodes présentées (notées \textit{concealer-1}, \textit{concealer-2} et \textit{concealer-3} dans la suite), plusieurs options ont été considérées pour la construction du signal \textit{concealer}. Il reste alors à reconstruire le signal \textit{concealer}, à le repasser dans le domaine temporel, puis à l'additionner avec la source. La reconstruction $C$ du signal \textit{concealer} dans le domaine fréquentiel se calcule comme suit :
$$C = m_C*e^{i*a_C}$$
L'application d'une TFCT inverse (toujours grâce à \textit{librosa}) sur $C$ permettra ensuite de passer le signal dans le domaine temporel, pour finalement effectuer l'addition $s + c$, représentant la mixture résultante de l'approche \textit{concealer}. 

\subsubsection{Choix de la méthode}\label{choix-methodes}
Afin d'évaluer une seule version de l'approche \textit{concealer}, il a fallu choisir l'une des méthodes de construction du signal \textit{concealer} présentées en section \ref{def-methodes}. Pour cela, une série de mixtures a été générée grâce à un protocole proposé par la bibliothèque \textit{doce}, disponible sous \textit{Python} et développée par Mathieu Lagrange \cite{lagrange2020doce}. Celle-ci permet de définir des facteurs, puis de générer toutes les combinaisons de facteurs grâce à une fonction appelée à chaque étape du protocole. Afin d'étudier les différentes méthodes de construction du signal \textit{concealer}, les facteurs suivants ont été définis : 
$$
\begin{cases}
    s = & [\textit{ventil1}] \\
    p = & [\textit{fountain, rain, stream, waterfall, waves}] \\
    method = & [\textit{concealer-1, concealer-2, concealer-3, masker}] \\
    msnr = & [\textit{0, -1}]
\end{cases}
$$

Ainsi, chaque mixture entre le bruit de ventilation enregistré à l'IRCAM et les sons d'eau sélectionnés a été générée pour toutes les méthodes (y compris l'approche \textit{masker}, représentée par le facteur "masker") et les deux valeurs de \textit{MSNR} calculés en dB LUFS \footnote{Les mixtures générées ainsi que les \textit{concealers} sont disponibles dans le dossier \textit{"sounds/method\_choice/"} et sont nommés de la manière suivante : \textit{approach=1+msnr=0\_fountain\_stimuli.wav}.}. \\

À l'écoute de ces mixtures, il est clair que la troisième méthode est meilleure que les méthodes \textit{concealer-1} et \textit{concealer-2}, pour lesquelles la ventilation apparaît beaucoup plus forte que le son d'eau. Ce dernier est en effet beaucoup plus audible avec la méthode \textit{concealer-3}, créant des mixtures qui n'augmentent pas le niveau perçu de la ventilation. Cette différence entre les méthodes est également audible à l'écoute des \textit{concealers} eux-mêmes. Pour les méthodes \textit{concealer-1} et \textit{concealer-2}, ceux-ci présentent une mixture entre son de ventilation et son d'eau, ce qui a pour effet d'ajouter une partie de la source lors de l'addition entre source et \textit{concealer} et donc de rendre celle-ci prédominante sur le son d'eau. Pour la méthode \textit{concealer-3} en revanche, le \textit{concealer} ne présente que le son d'eau, ce qui implique un ajout de son positif uniquement lors de la mixture entre source et \textit{concealer}. \\

Ce faisant, la méthode \textit{concealer-3} a été retenue pour construire les différents signaux \textit{concealers}. En plus de la comparaison entre méthodes, il est intéressant de comparer les mixtures créées avec la méthode \textit{concealer-3} avec celles obtenues avec l'approche \textit{masker}. En écoutant les différentes mixtures, il n'est apparu que très peu de différences entre les deux approches pour les niveaux étudiés, le seul contraste étant que le son d'eau paraissait plus brillant avec l'utilisation de l'approche \textit{concealer}. Cette première comparaison entre les deux approches a permis de valider le potentiel de l'approche \textit{concealer} face à l'approche \textit{masker} pour les \textit{MSNR} choisis.

\newpage
\section{Expérience perceptive}\label{expe}
Maintenant que les deux approches sont fonctionnelles, il est possible de créer des mixtures entre source et \textit{masker} (approche \textit{masker}) et entre source et \textit{concealer} (approche \textit{concealer}). La prochaine et dernière étape du travail mené lors de ce stage est de les évaluer. Une expérience perceptive a été conduite, et cette dernière partie en résumera les objectifs et présentera la méthodologie employée ainsi que les premiers résultats obtenus. 

\subsection{Présentation de l'expérience}\label{objectifs}
La première chose à définir pour une expérience perceptive est la dimension à évaluer. Si le projet \textit{ReNAR} vise à améliorer le confort, il a été souligné dans l'état de l'art (section \ref{etat-art}) que cette notion était difficile à évaluer. Une première option a donc été de tenter d'évaluer la gêne induite par les sources choisies. Cette notion de gêne semble d'autant plus pertinente à étudier dans le contexte d'un environnement de travail, car elle peut influer sur le bien-être mais aussi sur la performance des usagers du lieu. En s'appuyant sur le protocole de l'étude de Brocolini et coll. \cite{brocolini_effect_2016}, il aurait en effet été possible d'effectuer plusieurs mesures de performance, de charge de travail et de gêne perçue. En revanche, la définition d'une tâche de performance pertinente aurait nécessité plus ample préparation et le temps disposé pour réaliser l'expérience ne le permettait pas. De plus, une tâche de performance demande un temps d'exposition aux stimuli assez long, ce qui n'était pas vraiment compatible avec la diversité des paramètres prévus pour l'expérience, qui imposait un grand nombre de stimuli. Celle-ci aurait été trop longue, ou aurait nécessité une certaine organisation avec un grand nombre de participants ou un découpage en plusieurs sessions pour chaque participant. 

Cela étant dit, il a été décidé de s'orienter vers la notion d'agrément, qui peut être évaluée plus rapidement avec une seule mesure d'agrément perçu. Cette notion semble être un bon compromis pour une première expérience exploratoire sur le thème de l'approche \textit{concealer}. \\

Ainsi, l'objectif de l'expérience qui a été menée est de comparer les approches \textit{masker} et \textit{concealer} en termes d'agrément, afin d'évaluer le potentiel de l'approche \textit{concealer} dans le traitement du bruit au niveau perceptif au sein d'un environnement intérieur tel que les bureaux ouverts. Une hypothèse générale selon laquelle l'approche \textit{concealer} a des chances d'être meilleure que l'approche \textit{masker} en termes d'agrément a donc été posée. Le facteur expérimental principal étant les deux approches, des facteurs secondaires comme le niveau, le type de source ou le type de positif ont été étudiés, afin d'espérer identifier certaines conditions sur ces facteurs pour lesquelles les mixtures générées grâce à l'approche \textit{concealer} sont effectivement perçues comme plus agréables que celles générées par l'approche \textit{masker}.

\subsection{Méthodologie}\label{methodologie}
Les objectifs de l'expérience étant posés, cette section présentera son déroulement, du recrutement des participants aux méthodes mises en œuvre.

\subsubsection{Participants}
Un total de $30$ personnes a participé à cette expérience ($9$ femmes et $21$ hommes), avec une distribution d'âge entre 22 et 57 ans, l'âge moyen étant de 30 ans et l'âge médian de 25 ans. Un questionnaire rempli à la fin de l'expérience a révélé des problèmes auditifs chez trois participants, à savoir des acouphènes pour deux d'entre eux et des pertes auditives au-dessus de $8$kHz au niveau des deux oreilles pour le dernier. Les résultats de ces participants n'ont pas été écartés. Les participants recrutés travaillaient pour la plupart à l'IRCAM, au sein du laboratoire STMS, ou encore dans les services de production, d'informatique, ou des ressources humaines. Quelques personnes extérieures à l'IRCAM ont aussi pris part à l'expérience.

\subsubsection{Stimuli}
Les stimuli ont été construits selon quatre facteurs indépendants : l'approche utilisée, le niveau des mixtures générées, le type de source, et le type de positif. Comme mentionné plus haut, l'approche utilisée (\textit{masker} ou \textit{concealer}) représente le facteur principal de l'expérience. Il reste donc à choisir les valeurs que prendront les trois facteurs secondaires avant de construire les stimuli.

\begin{figure}[h!]
    \centering
    \includegraphics[scale = 0.7]{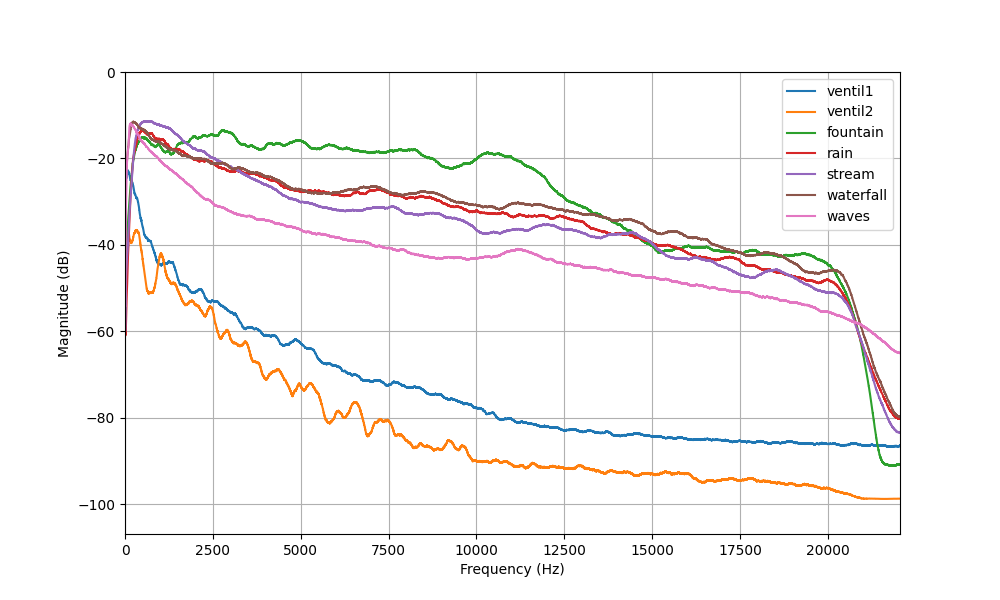}
    \caption{Spectres lissés des deux bruits de ventilation et des cinq sons d'eau pour une plage de fréquences allant de 0 à 20kHz}
    \label{spectres-global}
\end{figure}

\paragraph{Choix des types de source et de positif :}
~~\\
Ce sont les deux bruits de ventilation et les cinq sons d'eau sélectionnés en section \ref{choix-sons} qui ont permis la création des stimuli suivant les deux approches de masquage \footnote{Ces sons sont disponibles dans les dossiers \textit{sounds/ventilation\_noises/} et \textit{sounds/water\_sounds/}.}. Comme expliqué dans la section \ref{choix-sons}, les sons d'eau sont utilisés comme positifs pour les deux approches \textit{masker} et \textit{concealer}. Leur contenu spectral ainsi que ceux des deux sources sont présentés sur la figure \ref{spectres-global} suivante. 

\begin{figure}[h!]
    \centering
    \includegraphics[scale = 0.7]{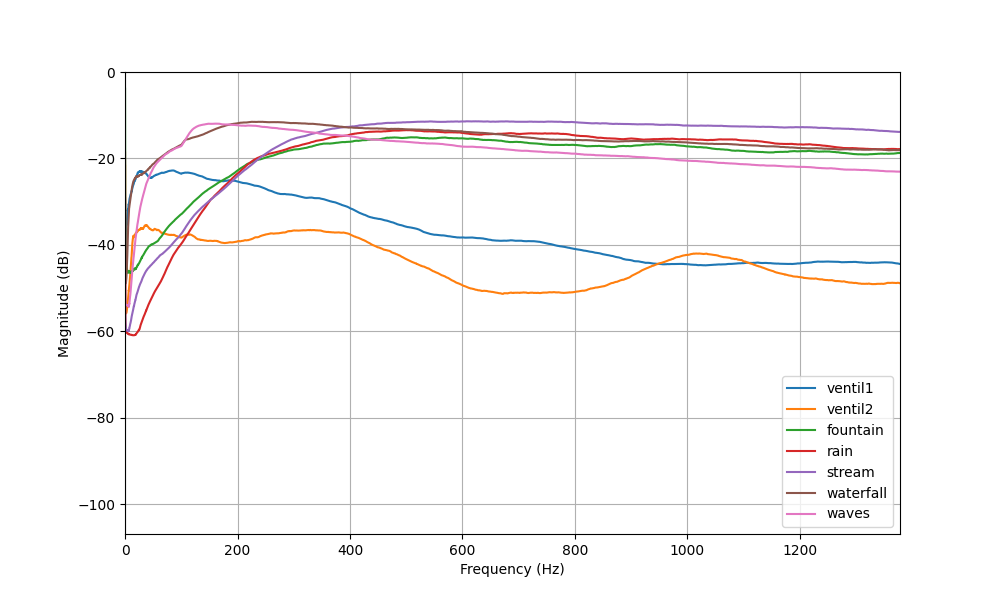}
    \caption{Spectres lissés des deux bruits de ventilation et des cinq sons d'eau pour une plage de fréquences allant de 0 à 1200Hz}
    \label{spectres-bf}
\end{figure}

Les bruits de ventilation présentent de l'énergie principalement dans les basses fréquences, alors que celle des sons d'eau décroît petit à petit jusqu'à $20kHz$. Pour détailler les endroits où tous les sons ont une énergie non négligeable, la figure \ref{spectres-bf} propose un zoom sur les basses fréquences.

\newpage

Il est maintenant plus facile d'observer que les sons \textit{fountain}, \textit{rain} et \textit{stream} n'ont que très peu d'énergie dans la plage couvrant les fréquences jusqu'à environ $200$Hz, contrairement aux deux bruits de ventilation, pour lesquels l'énergie décroît au-delà de cette fréquence. Restent les sons \textit{waterfall} et \textit{waves}, dont l'énergie dans les très basses fréquences dépasse celle des bruits de ventilation.

\paragraph{Choix du niveau des stimuli :}
~~\\
Le niveau des stimuli représente un facteur secondaire mais essentiel pour cette étude, car le projet \textit{ReNAR} repose sur un paradigme de minimisation de l'intensité sonore ajoutée. Il est alors nécessaire de s'assurer que le niveau de la mixture entre le bruit et le maquilleur ne se trouve pas nettement supérieur au niveau du bruit seul, et ce pour les deux approches.\\

Grâce aux premiers essais réalisés avec le paramètre \textit{MSNR} lors du choix de la méthode de construction du signal \textit{concealer} (section \ref{choix-methodes}), une réflexion s'est entamée sur la manière de contrôler le niveau non pas du maquilleur utilisé, mais plutôt de la mixture générée par les deux approches. Pour assurer directement un niveau des mixtures qui ne soit pas nettement supérieur à celui de la source, il a alors été décidé de raisonner selon un autre paramètre de niveau, noté $\Delta L_{Aeq}$, représentant la différence de niveau entre une mixture et la source utilisée pour cette mixture (en dB(A)). Contrairement au \textit{MSNR}, ce nouveau paramètre est positif ou nul (il n'est pas possible que le niveau d'une mixture soit inférieur à celui de la source puisque celle-ci est définie par l'addition entre la source et un maquilleur), et permet de contrôler directement le niveau de la mixture qui sera présentée comme stimuli aux participants de l'expérience perceptive. \\

Tout comme pour le choix de la méthode de construction du \textit{concealer}, un essai a été réalisé grâce au protocole proposé par la bibliothèque \textit{doce} \cite{lagrange2020doce} avec les facteurs suivants : 

$$
\begin{cases}
    s = & \mbox{[\textit{ventil1, ventil2}]} \\
    p = & \mbox{[\textit{fountain, rain, stream, waterfall, waves}]} \\
    method = & \mbox{[\textit{concealer-3, masker}]} \\
    \Delta L_{Aeq} = & \mbox{[\textit{0, 1, 2, 3, 4, 5, 6}]}
\end{cases}
$$

Pour cet essai, la source a été normalisée à $65$dB(A), car le niveau enregistré à $33$dB(A) a paru trop faible pour se placer dans le contexte d'une source sonore désagréable. En effet, les études menées sur le masquage ont présenté des niveaux beaucoup plus élevés (notamment celle de Cai et coll., qui utilisent un bruit de soudeuse électrique à $80$dB(A) \cite{cai_effect_2019}). Ainsi, un $\Delta L_{Aeq}$ de $0$ impose un niveau de $65$dB(A) pour les mixtures générées avec les deux approches. Plus précisément, il a fallu trouver le niveau du positif tel que le niveau de la mixture soit égal à $65\pm0.1$dB(A) \footnote{Une erreur maximale de $0.1$dB(A) a été autorisée sachant que le seuil de discrimination du niveau sonore est de l'ordre de $1$dB(A), donc bien supérieur à $0.1$dB(A).}, et ce pour les deux approches. \\

\begin{figure}[!h]
         \centering
         \begin{subfigure}[]{0.40\textwidth}
             \centering
             \includegraphics[scale = 0.6]{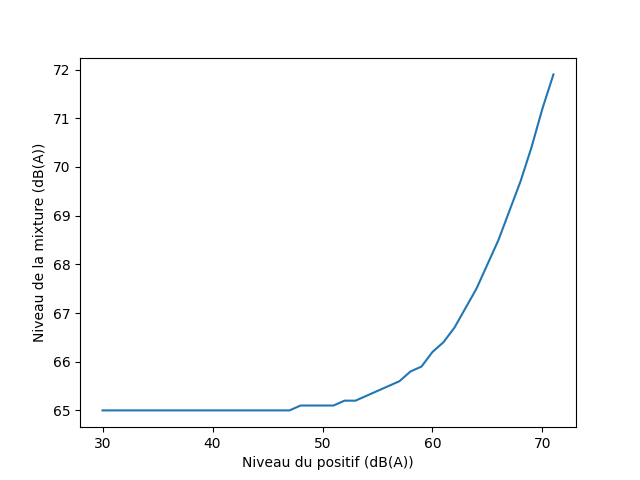}
             \caption{Pour l'approche masker}
         \end{subfigure}
         \hspace{50pt}
         \begin{subfigure}[]{0.40\textwidth}
             \centering
             \includegraphics[scale = 0.6]{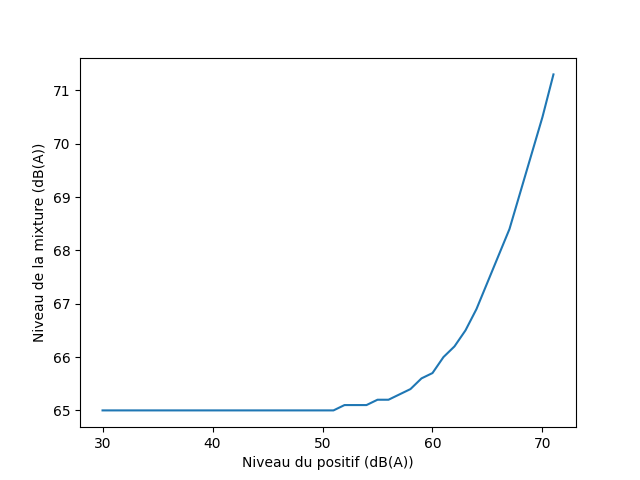}
             \caption{Pour l'approche concealer (méthode 3)}
         \end{subfigure}
        \caption{Niveau de la mixture en fonction du niveau du positif pour le bruit de ventilation enregistré à l'IRCAM et le son de fontaine}
        \label{levels}
\end{figure}

Cette relation entre niveau du positif et niveau de la mixture a été étudiée et sauvegardée pour chaque son d'eau, comme présenté sur la figure \ref{levels} dans le cas du son de fontaine.

Pour déterminer le niveau du positif tel que le niveau de la mixture soit égal à $65 + \Delta L_{Aeq} \pm 0.1$dB(A) (le niveau cible), une technique de \textit{Grid Search} a été utilisée. Ce type de technique est un processus qui effectue une recherche exhaustive à travers un sous-ensemble spécifié manuellement. Un ensemble de niveaux de positifs a donc été défini avant de lancer une recherche telle que : pour chaque niveau de cet ensemble, une mixture est calculée et son niveau est comparé au niveau cible dépendant de la valeur de $\Delta L_{Aeq}$. Si le niveau de la mixture est égal au niveau cible à $0.1$dB(A) près, alors la recherche s'arrête. 

Ainsi, le protocole \textit{doce} et la recherche de type \textit{Grid Search} ont permis d'une part de calculer les niveaux des positifs nécessaires à la création de mixtures dont le niveau $N_{mixture}$ est égal à $N_{source} + \Delta L_{Aeq}$, et ce pour les deux sources et les cinq positifs retenus. D'autre part, ces mixtures ainsi que les \textit{concealers} utilisés pour l'approche \textit{concealer} ont été générés et enregistrés au format \textit{wav}. \\

À l'écoute des mixtures générées \footnote{Celles-ci sont disponibles dans le dossier \textit{sounds/level\_choice/}, avec une nomenclature de la forme \textit{"approach=3+delta\_laeq=0+source=ventil1\_fountain\_stimulus.wav"}.}, il est intéressant de remarquer de grandes différences entre les valeurs de $\Delta L_{Aeq}$. Par exemple, le son d'eau n'est quasiment pas perceptible pour $\Delta L_{Aeq} = 0$dB(A), mais le devient assez clairement pour $\Delta L_{Aeq} = 1$dB(A), surtout pour les sons \textit{fountain} et \textit{rain}. Aussi, plus $\Delta L_{Aeq}$ augmente, plus les mixtures se différencient selon les deux approches. En effet, pour des valeurs de $\Delta L_{Aeq} \geq 4$dB(A), l'approche \textit{concealer} produit des mixtures pour lesquelles le son d'eau semble beaucoup plus brillant que pour l'approche \textit{masker}. Finalement, les niveaux obtenus pour les mixtures étant considérés comme trop forts pour $\Delta L_{Aeq} \geq 4$dB(A), il a été décidé de garder des valeurs de $\Delta L_{Aeq}$ allant de $0$ à $3$dB(A), par pas de $0.5$dB(A) afin d'affiner la perception du positif.

\paragraph{Construction des stimuli :}
~~\\
Les facteurs de l'expérience ainsi fixés, les stimuli ont pu être construits. Le niveau sonore équivalent pondéré A des sources de bruit a été fixé à $65$dB(A), et la construction des stimuli a été entreprise grâce aux fonctionnalités de la bibliothèque \textit{doce} \cite{lagrange2020doce} sous \textit{Python}, qui ont permis de générer des mixtures entre les sources et les maquilleurs pour les deux approches. La moitié des stimuli a donc été construite selon l'approche \textit{masker}, et l'autre moité selon l'approche \textit{concealer}. 

De plus, au cours de la génération des mixtures, ces dernières ont été déclinées en plusieurs niveaux selon le paramètre $\Delta L_{Aeq}$. Une valeur de $\Delta L_{Aeq} = 2.5$dB(A) signifie donc que le niveau de la mixture générée est fixé à $67.5$dB(A), soit $2.5$dB(A) de plus que la source seule. Chaque paire constituée d'une source et d'un maquilleur a ainsi donné sept mixtures différentes, créant des stimuli de $10$s enregistrés au format \textit{wav} avec une fréquence d'échantillonnage de $44.1$kHz. \\

Au final, un total de $140$ stimuli a été généré selon les facteurs suivants :
$$
\begin{cases}
    s = & \mbox{[\textit{ventil1, ventil2}]} \\
    p = & \mbox{[\textit{fountain, rain, stream, waterfall, waves}]} \\
    method = & \mbox{[\textit{concealer-3, masker}]} \\
    \Delta L_{Aeq} = & \mbox{[\textit{0, 0.5, 1, 1.5, 2, 2.5, 3}]}
\end{cases}
$$

\subsubsection{Apparatus}
L'expérience a eu lieu dans une cabine audiométrique dans les locaux de l'équipe Perception et Design Sonores à l'IRCAM. Les stimuli ont été présentés grâce à un casque ouvert \textit{Beyerdynamic DT 990 Pro} avec une impédance de $250\Omega$, connecté à un \textit{MacMini}. Avant d'accueillir les participants, le casque a été calibré grâce à un sonomètre. Le réglage du niveau a été effectué à l'aide des boutons de volume de l'ordinateur, avec un volume retenu de $7$ points, de telle sorte que les deux sources normalisées à $65$dB(A) soient à un niveau d'environ $50$dB(A) dans le casque (un niveau supérieur ayant été jugé trop élevé pour les participants). 

\subsubsection{Protocole}
Avant de prendre place dans la cabine, chaque participant a pris connaissance du déroulement de l'expérience (fiche "Consigne" en Annexe \ref{consigne}), et a signé une "Feuille de Consentement" (également disponible en Annexe \ref{consentement}). Une fois installés dans la cabine, il a bien été spécifié aux participants de ne pas toucher au volume de l'ordinateur. L'expérience leur a ensuite été présentée grâce à une interface web codée en \textit{JavaScript} à l'aide de la bibliothèque \textit{jsPsych} \cite{deleeuw2023jspsych}. \\

\begin{figure}[h!]
    \centering
    \includegraphics[scale = 0.35]{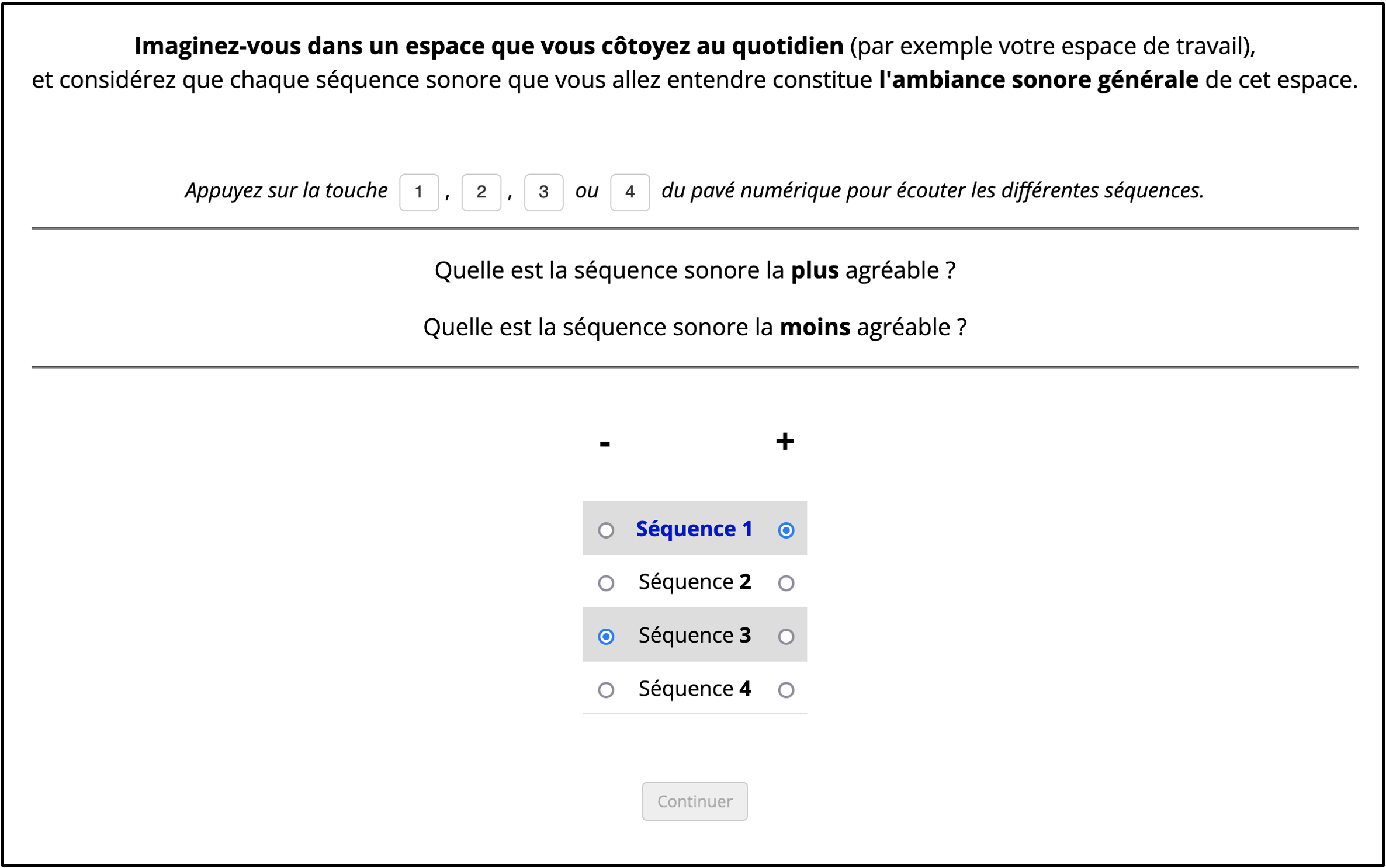}
    \caption{Interface présentée aux participants lors de la première partie de l'expérience, visant à juger des groupes de quatre séquences en termes d'agrément}
    \label{partie1-expe}
\end{figure}

Dans un premier temps, la partie principale de l'expérience faisant appel à la méthode \textit{Best-Worst Scaling (BWS)} a été présentée. Cette première partie d'expérience consistait à demander aux participants d'évaluer des groupes de $4$ "séquences sonores" (appelés 4-tuples dans la suite) en sélectionnant celle qu'ils jugeaient comme la plus agréable et celle qu'ils jugeaient comme la moins agréable. Ce type de protocole, ainsi que la manière d'en analyser les résultats, a fortement été inspiré de l'étude menée par Matthieu Duroyon dans le cadre de sa thèse sur le confort sonore au sein des véhicules électriques \cite{duroyon2025confort}. La figure \ref{partie1-expe} présente l'interface proposée aux participants pour effectuer leur jugement sur un 4-tuple. Plus précisément, la première partie s'est déroulée en trois temps : \\

- \textbf{Phase d'entraînement} : deux premiers 4-tuples ont d'abord été présentés pour permettre aux participants de se familiariser avec l'interface et les stimuli (les données n'ont pas été enregistrées pour ces deux 4-tuples, qui étaient les mêmes pour tous les participants).

- \textbf{Phase principale} : $35$ 4-tuples ont suivi cette phase d'entraînement, constituant le coeur de l'expérience. Ceux-ci ont été constitués avec les $140$ stimuli, de sorte que chaque participant entende une seule fois tous les stimuli, et que les 4-tuples qui lui sont présentés soient uniques. Ces deux contraintes ont été assurées grâce à un script \textit{Python} fourni par Matthieu Duroyon, qui a permis de construire les groupes de stimuli de chaque participant avant la passation.

- \textbf{Phase de \textit{re-test}} : enfin, quatre 4-tuples supplémentaires ont été présentés pour vérifier la cohérence intra-participants. Ceux-ci ont pour chaque participant été sélectionnés de manière aléatoire parmi les 4-tuples présentés durant la phase principale, et n'ont pas été séparés des autres afin de ne pas prévenir les participants de cette nouvelle phase. 

\newpage

Comme présenté sur la figure \ref{partie1-expe} ci-dessus, une mise en contexte a aussi été implémentée. Ce choix a résulté du test pilote réalisé pour vérifier le bon fonctionnement de l'expérience. À ce moment là, l'absence de contexte et l'utilisation du mot "Son" à la place de "Séquence" a paru manquer à l'expérience. En effet, le fait de juger simplement des sons de $10$s ou d'imaginer que ces sons constituent l'ambiance sonore d'un lieu ne conditionne pas du tout le même état d'esprit. Il a donc été jugé nécessaire de formuler une légère mise en contexte qui oriente les participants vers leurs propres espaces de travail. \\

\begin{figure}[h!]
    \centering
    \includegraphics[scale = 0.6]{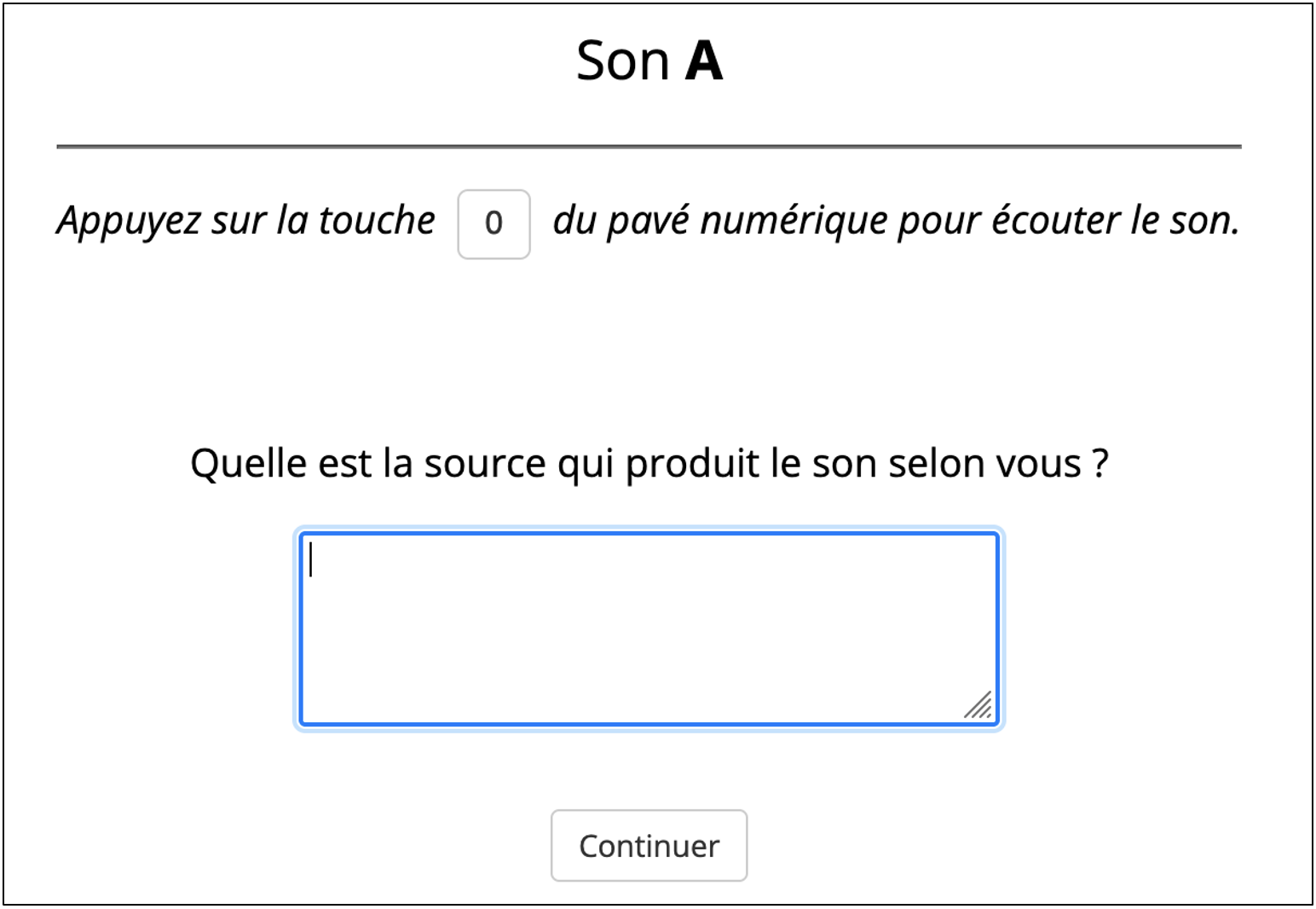}
    \caption{Interface présentée aux participants lors de la seconde partie de l'expérience, visant à étudier la reconnaissance de source des différents positifs}
    \label{partie2-expe}
\end{figure}

À la suite de la première partie de l'expérience, une seconde partie a été développée autour d'une tâche de verbalisation. Au vu des différents types de positifs (les cinq sons d'eau) utilisés dans la création des stimuli, il a en effet paru intéressant de relever la proportion de participants qui ont reconnu les sources sonores produisant ces sons d'eau. Au cours de la seconde partie, les cinq positifs ont alors été présentés aux participants, et il leur a été demandé de décrire la source du son proposé, comme exposé sur la figure \ref{partie2-expe} ci-dessous. Les cinq sons ont été présentés l'un après l'autre, et dans un ordre aléatoire. 

La seconde partie concluant l'expérience, la récolte des données se déclenche grâce au bouton "Terminer l'expérience" présenté sur l'interface. Toutes les réponses des participants pour les deux parties ont ainsi été sauvegardées au format \textit{csv} dans des fichiers nommés \textit{"results(numéro du participant).csv"}. La durée de l'expérience, estimée à $30$min, a également été enregistrée pour chaque participant. \\

Pour terminer, il a été demandé aux participants de compléter le questionnaire présenté en Annexe \ref{questionnaire}, afin de recueillir quelques informations supplémentaires sur leur point de vue et leur ressenti par rapport au problème de bruit, et notamment sur leur lieu de travail. Quelques remarques intéressantes de certains participants seront présentées dans la section  \ref{resultats} suivante.

\subsection{Résultats}\label{resultats}
Les deux parties de l'expérience ont permis de fournir plusieurs types de résultats, qui seront présentés dans cette partie, et grâce auxquels les hypothèses de départ ont été confirmées ou réfutées.

\subsubsection{Durée de l'expérience}
Tout d'abord, la durée de l'expérience a été récoltée pour chaque participant, afin de s'assurer que l'estimation de 30min était bonne, et pour tenter de déceler des personnes très rapides ou très longues par rapport à l'ensemble des participants, ces personnes pouvant être reconnues comme \textit{outliers}. Cette durée correspond plus précisément au temps écoulé entre le moment où le participant appuie sur le bouton "Commencer l'expérience" et le moment où il appuie sur le bouton "Terminer l'expérience", tous deux présentés sur l'interface graphique permettant le déroulement de l'expérience. La durée moyenne calculée grâce au passage des $30$ participants s'élève à 36min (durée arrondie à l'unité). La durée la plus courte enregistrée est de 16min et la plus longue de 66min. La médiane étant de 35min, l'estimation d'une durée d'expérience de 30min semble néanmoins correcte, et aucun \textit{outlier} n'a été détecté.

\subsubsection{Choix de l'algorithme de notation}
La première partie de l'expérience a fait appel à la méthode \textit{BWS}, et les jugements pour chaque 4-tuples ont été enregistrés pour chaque participant. Après un formatage des données, un script \textit{Python} a permis de calculer les scores de chaque stimuli, et de les classer en termes d'agrément. Les scores ont été calculés grâce à l'algorithme \textit{Value Learning} présenté en section \ref{BWS} de l'état de l'art. Cet algorithme a été retenu sur la base des résultats des simulations menées par Hollis pour tester la performance de plusieurs algorithmes de notation \cite{hollis_scoring_2018}. En effet, même si les algorithmes Elo et Rescorla-Wagner sont performants, c'est \textit{Value Learning} qui s'est trouvé le moins affecté par le bruit du à la variabilité inter et intra-participants. Puisque l'expérience menée dans le cadre de cette étude implique des jugements humains à l'origine de bruit dans les données, cet algorithme a semblé le plus adapté pour calculer les scores des stimuli sonores utilisés.

\subsubsection{Cohérence inter et intra-participants}
\begin{figure}[h!]
    \centering
    \includegraphics[scale = 0.62]{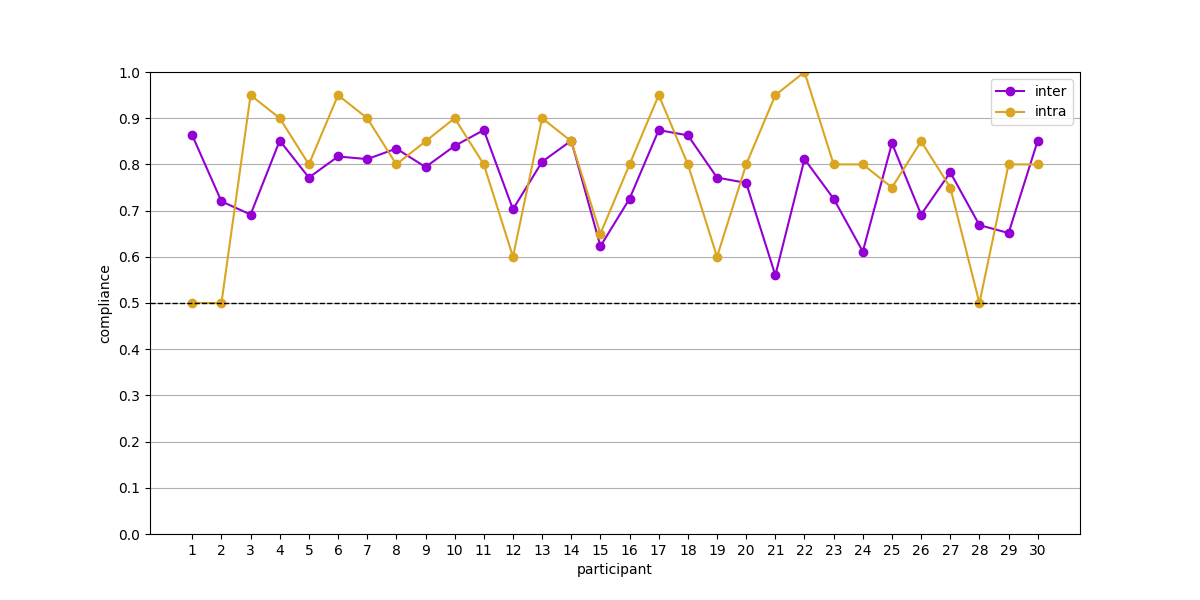}
    \caption{Cohérence inter-participants (en bleu) et intra-participants (en orange) pour chaque participant}
    \label{compliance}
\end{figure}

Grâce à des fonctions de conformité, des valeurs de cohérence ont été calculées, afin de savoir à quel point chaque participant est cohérent avec les autres et avec lui-même. Le principe de ces fonctions, implémentées en \textit{Python}, est de vérifier pour chaque 4-tuple si les relations entre les sons suggérées par le choix du meilleur et du pire se retrouvent bien en termes de score. En effet, lorsque qu'un participant choisit un meilleur et un pire parmi les quatre stimuli qui lui sont présentés, cinq relations peuvent être déduites, comme expliqué dans la section \ref{BWS}. Pour chaque participant et chaque 4-tuple qui lui a été proposé, il est possible de récupérer ces cinq relations, et de vérifier si elle est respectée en termes de scores. En comparant les relations choisies par un participant aux scores déduits de l'ensemble de l'expérience (avec tous les participants), il est ainsi possible de calculer une cohérence inter-participants. Si ces relations sont comparées aux scores déduits de ce même participant, il est possible d'obtenir une cohérence intra-participants. Cela étant dit, la figure \ref{compliance} ci-dessous montre les résultats obtenus.

\newpage

La ligne en pointillés marquant la valeur 0.5 décrit la valeur minimale que peuvent prendre les valeurs de cohérence. Ce minimum pourrait être assimilé au "hasard", puisqu'il suggère que la moitié des jugements d'un participant ne correspondent pas aux scores obtenus pour l'ensemble des participants, ou pour ce même participant. En observant la figure \ref{compliance}, la cohérence inter-participants est supérieure à $0.7$ pour la plupart des participants, mis à part les participants $15$, $21$, $24$, $28$ et $29$ (seul le participant $21$ a obtenu une valeur inférieure à $0.6$). Par rapport à la valeur minimale $0.5$, les participants sont donc assez cohérents entre eux. Concernant la cohérence intra-participants, elle est également assez élevée pour la plupart des participants, dépassant souvent $0.8$. Cela signifie que ces participants sont très cohérents avec eux-même, à l'inverse des participants $1$, $2$ et $28$, qui obtiennent des valeurs de $0.5$. Ces résultats assez bas étaient attendus car certains participants ont signifié à la fin de l'expérience qu'il avait été difficile pour eux de savoir ce qui leur plaisait réellement, et que les réponses données au début de l'expérience ne seraient sûrement pas cohérentes avec les réponses données à la fin de celle-ci.

\subsubsection{Effet de l'approche sur le score}
\begin{figure}[h!]
    \centering
    \includegraphics[scale = 0.7]{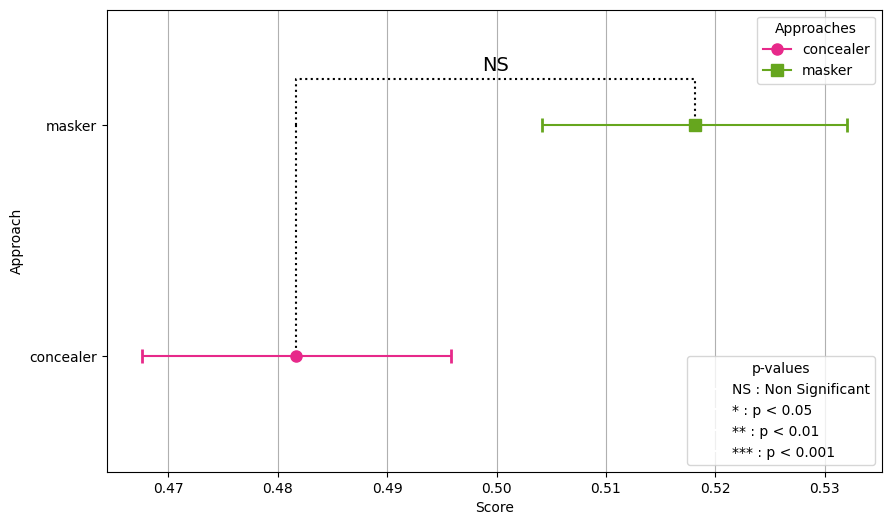}
    \caption{Scores moyens des mixtures en fonction de l'approche utilisée (concealer en rond rose et masker en carré vert)}
    \label{errorbar_approach}
\end{figure}

Une fois les scores calculés pour tous les stimuli, il est possible de les visualiser en fonction des paramètres de l'expérience. Pour tenter de répondre à la problématique principale, les données ont tout d'abord été agrégées selon les deux approches. La figure \ref{errorbar_approach} ci-dessous montre néanmoins que la différence entre les scores obtenus avec les approches \textit{masker} et \textit{concealer} n'est pas significative. Il n'est donc pas possible à ce stade de prétendre que l'approche \textit{concealer} donne de meilleurs résultats que l'approche \textit{masker}, la figure \ref{errorbar_approach} suggérant l'inverse.

\newpage

L'approche \textit{masker} est globalement meilleure que l'approche \textit{concealer} en termes de score, et l'utilisation d'un \textit{t-test} fournissant une \textit{p-value} $p = 0.07$ montre que la différence des scores moyens observés pour les deux approches n'est pas significative. Les participants ont donc globalement jugé les mixtures générées grâce à l'approche \textit{masker} comme plus agréables que celles générées grâce à l'approche \textit{concealer}.

\subsubsection{Effet de la source sur le score}
Afin de comprendre les effets des paramètres secondaires sur les scores et d'espérer trouver certains paramètres pour lesquels l'approche \textit{concealer} fournit de meilleurs résultats que l'approche \textit{masker}, il est possible de séparer les données obtenues selon les différents paramètres.

\begin{figure}[h!]
    \centering
    \includegraphics[scale = 0.7]{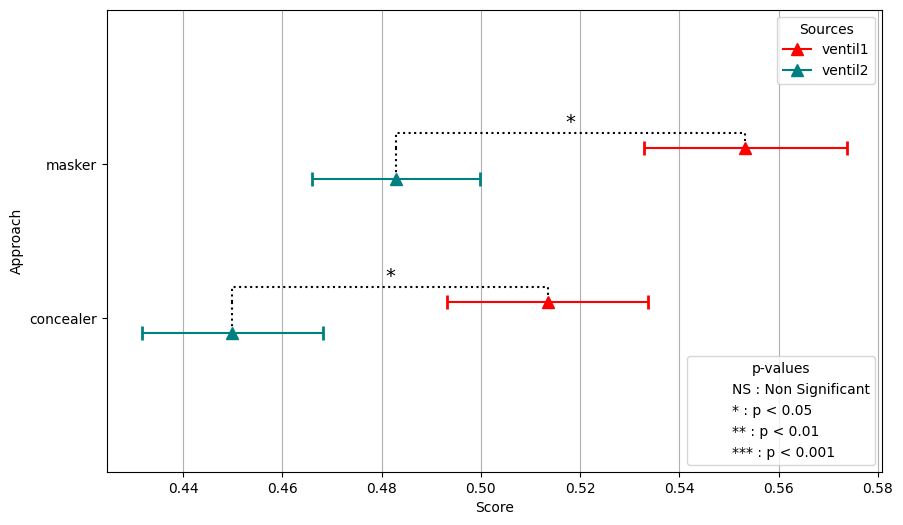}
    \caption{Scores moyens des mixtures en fonction de l'approche utilisée et des deux sources (ventil1 en rouge et ventil2 en bleu)}
    \label{errorbar_sources}
\end{figure}

En premier lieu, il est pertinent d'étudier l'effet du type de source sur le score. La figure \ref{errorbar_sources} suivante présente les scores moyens recueillis pour les deux sources selon les deux approches. 

\newpage

Cette figure et les \textit{t-tests} réalisés montrent que la différence entre les deux sources est significative, et ce pour les deux approches. La source \textit{ventil1}, qui représente le bruit de ventilation enregistré à l'IRCAM, est mieux perçu que la source \textit{ventil2}. Ce résultat était attendu puisque la source \textit{ventil2} a été choisi pour son caractère plus gênant dû à sa composante tonale. Aussi, puisque les deux sources ont des effets différents sur le score, et donc sur l'agrément, les résultats les concernant ne seront pas agrégés dans la suite. Une analyse en parallèle sera donc effectuée pour séparer les scores des mixtures contenant \textit{ventil1}, et ceux des mixtures contenant \textit{ventil2}.

\subsubsection{Effet du niveau sur le score}
\begin{figure}[!h]
         \centering
         \begin{subfigure}[]{0.70\textwidth}
             \centering
             \includegraphics[width=\textwidth]{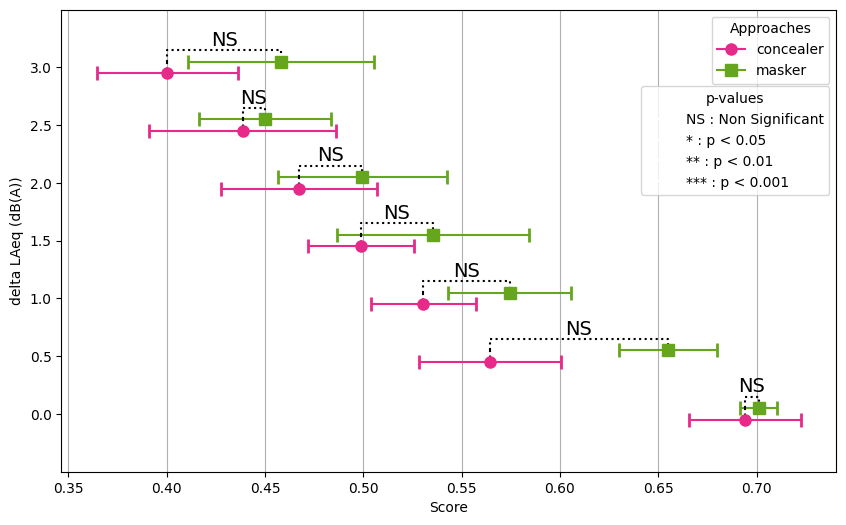}
             \caption{Scores pour "ventil1"}
             \label{level_ventil1}
         \end{subfigure}
         \begin{subfigure}[]{0.70\textwidth}
             \centering
             \includegraphics[width=\textwidth]{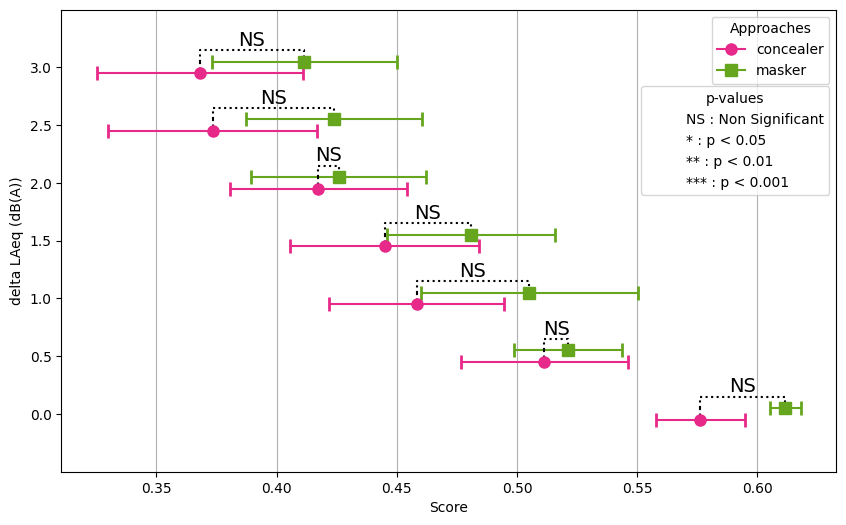}
             \caption{Scores pour "ventil2"}
             \label{level_ventil2}
         \end{subfigure}
        \caption{Scores moyens des mixtures en fonction de $\Delta L_{Aeq}$ pour les deux approches masker (carré vert) et concealer (rond rose)}
        \label{errorbar_levels}
\end{figure}

L'effet du niveau sur le score est considérable, et la figure \ref{errorbar_levels} montre une tendance linéaire entre le score et le paramètre $\Delta L_{Aeq}$, et ce indépendamment de la source utilisée pour la génération des mixtures. 

\newpage

Les figures \ref{level_ventil1} et \ref{level_ventil2} montrent une tendance décroissante du score lorsque le paramètre $\Delta L_{Aeq}$ augmente, et suggèrent une relation presque linéaire. Pour les deux sources de bruit, ce sont les mixtures générées pour $\Delta L_{Aeq} = 0$dB(A) qui sont les mieux perçues.

De plus, l'approche \textit{masker} reste meilleure que l'approche \textit{concealer} pour toutes les valeurs de $\Delta L_{Aeq}$ en termes de score, et donc en termes d'agrément. Il arrive même que l'approche \textit{masker} fournisse de meilleurs scores que l'approche \textit{concealer} pour des valeurs plus élevées de $\Delta L_{Aeq}$. Par exemple, les mixtures créées grâce à l'approche \textit{masker} obtiennent de meilleurs scores pour $\Delta L_{Aeq} = 1.5$dB(A) que celles créées à l'aide de l'approche \textit{concealer} pour $\Delta L_{Aeq} = 1$dB(A). Cependant, la différence entre les deux approches n'est toujours pas significative, et ceux pour les deux sources de bruit. 

\subsubsection{Effet du positif sur le score}
\begin{figure}[!h]
         \centering
         \begin{subfigure}[]{0.70\textwidth}
             \centering
             \includegraphics[width=\textwidth]{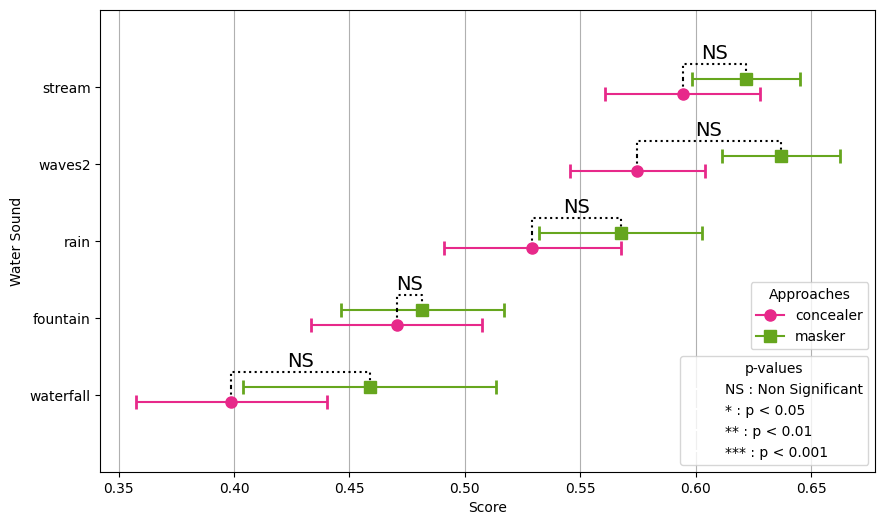}
             \caption{Scores pour "ventil1"}
             \label{watersound_ventil1}
         \end{subfigure}
         \begin{subfigure}[]{0.70\textwidth}
             \centering
             \includegraphics[width=\textwidth]{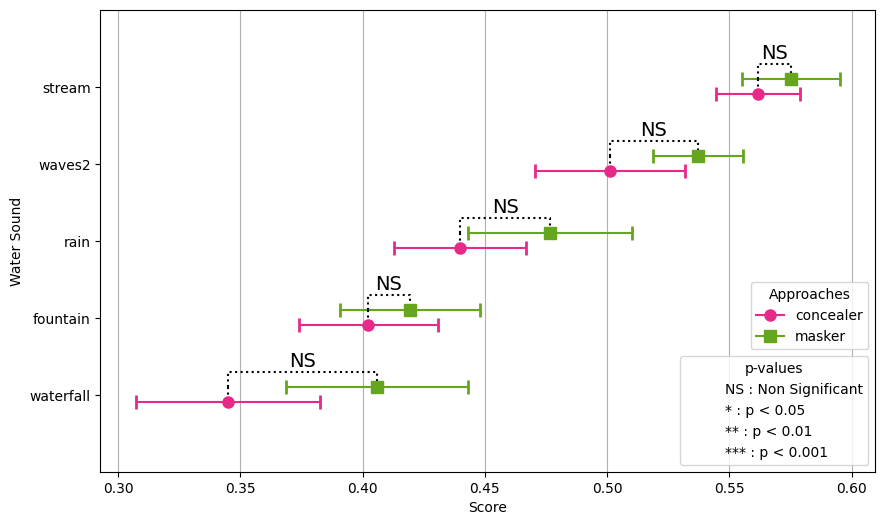}
             \caption{Scores pour "ventil2"}
             \label{watersound_ventil2}
         \end{subfigure}
        \caption{Scores moyens des mixtures en fonction du type de son d'eau pour les deux approches masker (carré vert) et concealer (rond rose)}
        \label{errorbar_watersounds}
\end{figure}

Le dernier paramètre, le type de positif, semble lui aussi avoir un effet significatif sur le score. La figure \ref{errorbar_watersounds} fait émerger un ordre parmi les positifs, plaçant les mixtures générées avec le son \textit{stream} comme les mieux perçues, tandis que celles générées avec le son \textit{waterfall} sont les moins bien perçues.

\newpage

Tout comme pour le paramètre de niveau des mixtures $\Delta L_{Aeq}$, les deux figures \ref{watersound_ventil1} et \ref{watersound_ventil2} confirment que pour chaque positif, l'approche \textit{masker} fournit de meilleurs résultats en termes de score que l'approche \textit{concealer}. Aucune différence significative n'est relevée entre les scores moyens obtenus pour chaque son d'eau avec les deux approches.

\subsubsection{Visualisation des paramètres de l'expérience}
\begin{figure}[!h]
         \centering
         \begin{subfigure}[]{0.90\textwidth}
             \centering
             \includegraphics[width=\textwidth]{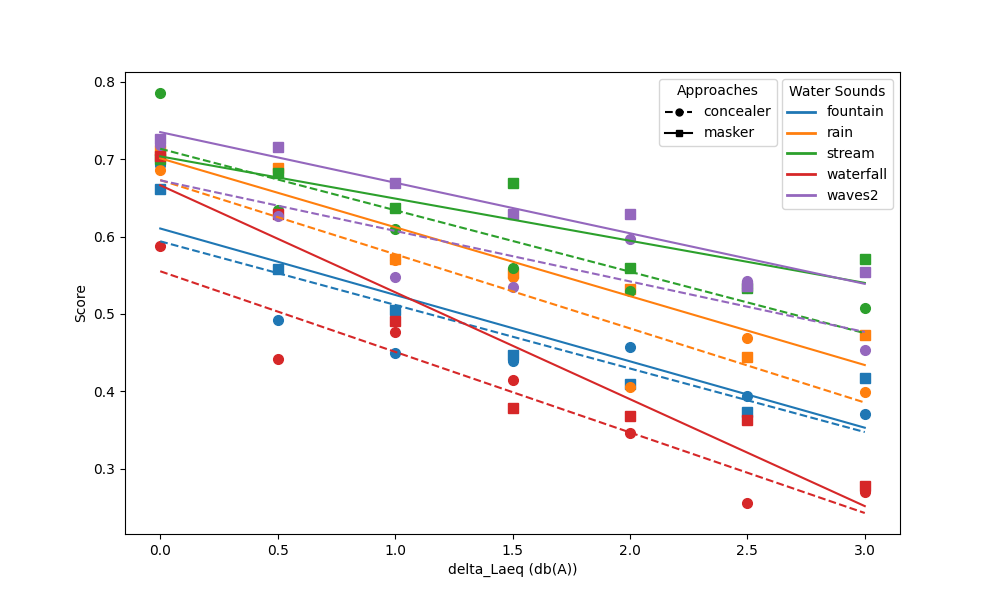}
             \caption{Scores pour "ventil1"}
             \label{allresults_ventil1}
         \end{subfigure}
         \begin{subfigure}[]{0.90\textwidth}
             \centering
             \includegraphics[width=\textwidth]{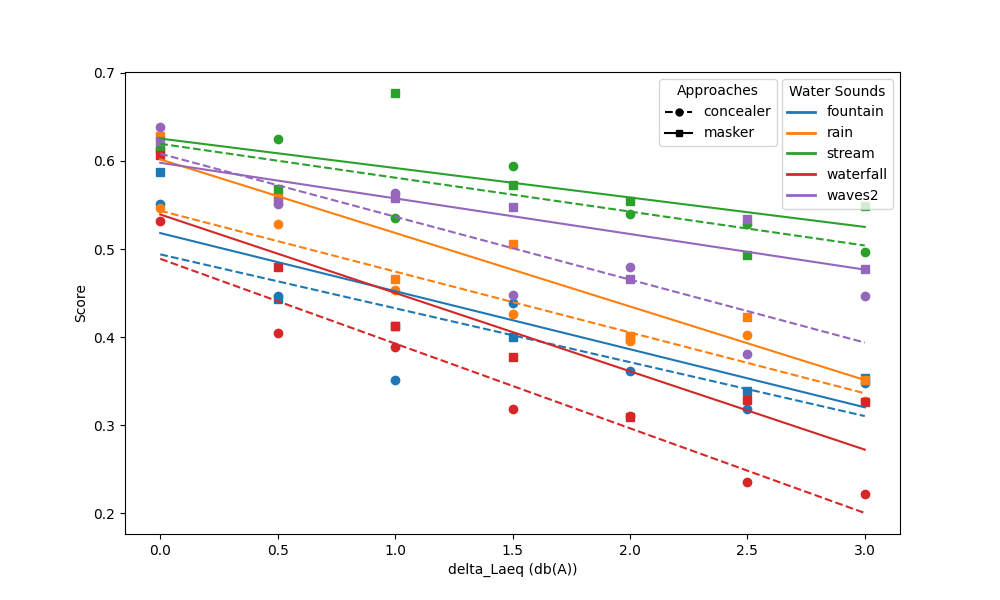}
             \caption{Scores pour "ventil2"}
             \label{allresults_ventil2}
         \end{subfigure}
        \caption{Scores des mixtures en fonction de $\Delta L_{Aeq}$, de l'approche (ronds et carrés) et du positif utilisé (bleu, orange, vert, rouge et violet) pour les deux bruits de ventilation}
        \label{allresults}
\end{figure}

Les figures précédentes, présentées sous forme d'\textit{error bars}, ont principalement appuyé le fait que la différence entre les deux approches n'est pas significative, et ce pour tous les paramètres de l'expérience. Seul l'effet de la source sur le score a pu être démontré par l'utilisation d'un \textit{t-test} puisque seulement deux sources ont été utilisées pour la génération des mixtures. Pour le type de positif et le niveau des mixtures, qui présentent un plus grand nombre de conditions, il n'a pas été possible d'employer le même type de test statistique, et le format des \textit{error bars} semble moins pertinent. C'est en ce sens que la figure \ref{allresults} ci-après présente tous les paramètres utilisés, et permet de se rendre compte de l'effet du paramètre $\Delta L_{Aeq}$ et du type de positif sur le score, et donc sur l'agrément. \\

Sur cette figure \ref{allresults}, les valeurs des scores des différentes mixtures sont représentées par les points, et des régressions linéaires (représentées par les droites) ont été calculées pour chaque approche et chaque type de positif. Cette visualisation permet dans un premier temps d'observer une tendance générale sur le niveau des mixtures, et de supposer une relation linéaire entre le score et le paramètre $\Delta L_{Aeq}$. En effet, comme spécifié plus haut, le score décroît lorsque la valeur de $\Delta L_{Aeq}$ augmente. Il est alors intéressant de remarquer que c'est une augmentation du niveau du positif qui implique une augmentation de $\Delta L_{Aeq}$ (puisque le niveau de la source est constant). Ces résultats traduisent donc le fait que l'ajout de positif à la source n'est pas apprécié, que ce soit avec l'utilisation de l'approche \textit{masker} ou de l'approche \textit{concealer}. Les participants semblent préférer les mixtures telles que $\Delta L_{Aeq} = 0$dB(A), qui sont perceptivement proches des deux sources. De la même manière, les décroissances de score lorsque $\Delta L_{Aeq}$ augmente prouvent que plus la part de son d'eau est forte dans les mixtures, moins ces dernières sont jugées agréables, et ce pour tous les types de sons d'eau. 

\newpage

D'autre part, cette figure suggère un classement des mixtures selon le positif utilisé. En notant $S_{positif}$ le score moyen des mixtures créées à l'aide du son d'eau indiqué, la relation suivante peut être supposée : 
$$S_{stream} > S_{waves} > S_{rain} > S_{fountain} > S_{waterfall}$$

Pour tenter de comprendre cette relation, plusieurs informations ont été récoltées sur les différents sons d'eau. Plusieurs indices spectraux ont d'abord été étudiés. Plus précisément, un indice d'entropie spectrale a été calculé grâce à la bibliothèque \textit{scikit-maad} \cite{ulloa2021scikit} sous \textit{Python}. La bibliothèque \textit{librosa} \cite{mcfee2015librosa} a également permis de calculer le centroïde spectral, la bande passante, le contraste spectral et la planéité spectrale de chaque son d'eau. La figure \ref{features} ci-dessous présente les valeurs de ces indices ainsi que les scores obtenus pour chaque son d'eau : 

\begin{figure}[h!]
    \centering
    \includegraphics[scale = 0.6]{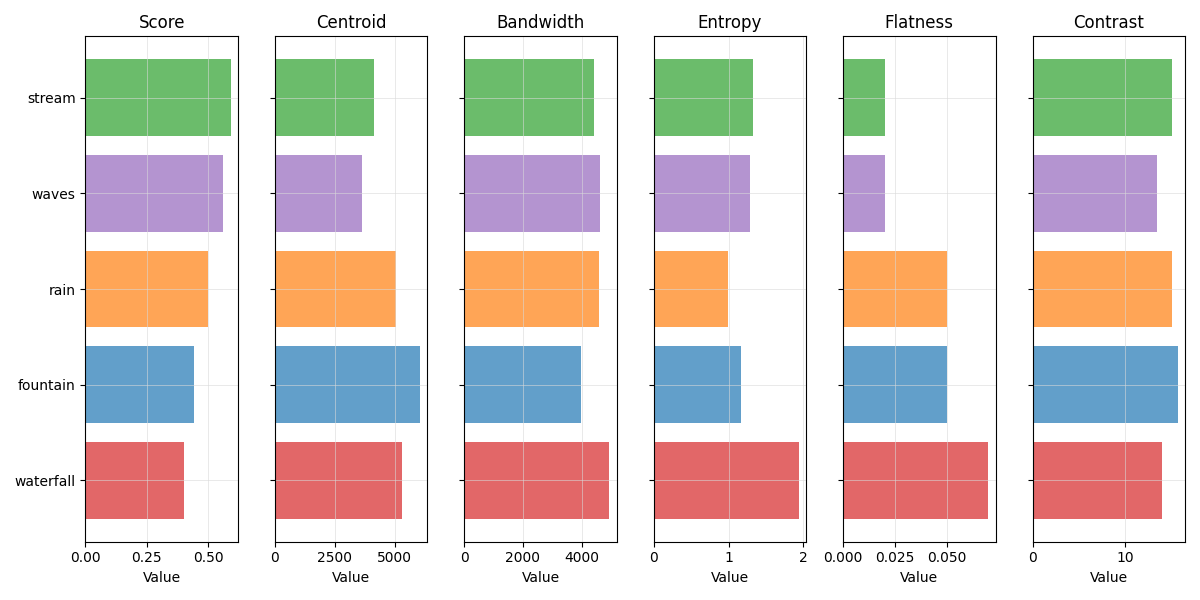}
    \caption{Scores moyens des mixtures et indices spectraux pour les cinq sons d'eau}
    \label{features}
\end{figure}

Cette figure \ref{features} permet d'observer les variations des indices calculés en fonction du type de son d'eau. En particulier, les variations de l'indice de planéité spectrale (\textit{Flatness} sur la figure) semble suivre une tendance inverse à celle du score. Sachant que les scores des mixtures générées à l'aide des différents sons d'eau représentent l'agrément perçu, et que la planéité spectrale quantifie la ressemblance d’un son avec le bruit par opposition à sa tonalité, une relation entre agrément et diversité spectrale peut être supposée. Les mixtures les mieux perçues sont celles constituées des sons \textit{stream} et \textit{waves}, qui ont les valeurs les plus faibles de planéité spectrale, et donc les moins proches spectralement du bruit blanc (c'est-à-dire les plus tonaux). À l'inverse, le son \textit{waterfall} présente une valeur élevée de planéité spectrale, et se rapproche donc du bruit blanc. Ce sont les mixtures contenant ce son qui sont les moins bien perçues. \\

En plus des valeurs physiques que fournissent les indices spectraux, la deuxième partie de l'expérience a permis de recueillir des informations sur la capacité des participants à reconnaître la source des sons d'eau. Les réponses des participants ont été étudiées et un pourcentage de reconnaissance a été initialisé pour chaque son d'eau. Si le participant définit clairement la source du son en mentionnant les noms des sons d'eau (fontaine, pluie, cours d'eau, cascade et vagues), le score de la source reconnue est augmenté de $1$. Pour les sons \textit{stream} et \textit{waves}, d'autres mots ont été acceptés, tant que le participant utilise bien ces mots pour décrire la source du son, et pas seulement un décor. Par exemple, les mots "rivière", "ruisseau", "torrent" et "fleuve" sont acceptés pour parler du son \textit{stream}, et le mot "mer" est accepté pour parler du son \textit{waves}. En revanche, "le vent dans les arbres à l'arrière d'une plage" n'est pas accepté, car la réponse suggère que la source du son est le vent, et non les vagues. Enfin, plusieurs participants ont ajouté du contexte dans leurs réponses, en donnant des descriptions assez précises de l'environnement qu'ils ont imaginé autour de la source du son. Cela ne changeant pas la réponse principale sur la source des sons, tous les contextes ont été acceptés. Ainsi, "pluie sur un toit" est autant accepté que "la pluie qui tombe et qui tape sur les carreaux", ou simplement "pluie", et "les vagues de la mer par temps calme" est autant accepté que "l'Atlantique par gros temps". Finalement, cette partie a révélé que le son \textit{rain} est le plus identifiable, suivi des sons \textit{stream} et \textit{waves}, suivis du son \textit{waterfall}, et enfin du son \textit{fountain}, que seulement quatre participants ont reconnu. Les résultats ont été relevés grâce à des pourcentages de reconnaissance des sources, présentés dans le tableau \ref{tab:reconnaissance_source} suivant : 
\begin{table}[h!]
\centering
\begin{tabular}{|c|c|c|c|c|c|}
\hline
\multirow{2}{*}{\vphantom{T}} & \multicolumn{5}{c|}{\textbf{Type de son d'eau}} \\ \cline{2-6} 
 & \textit{fountain} & \textit{rain} & \textit{stream} & \textit{waterfall} & \textit{waves} \\ \hline
\begin{tabular}[c]{@{}c@{}}\textbf{Reconnaissance de}\\ \textbf{la source (en \%)}\end{tabular} &  13.3 & 83.3 & 60 & 26.7 & 60 \\ \hline
\end{tabular}
\caption{Reconnaissance de la source des sons d'eau utilisés pour la création des mixtures}
\label{tab:reconnaissance_source}
\end{table}

La reconnaissance des sons d'eau ne semble pas être le facteur principal responsable de l'agrément perçu dans le cadre des mixtures créées avec les deux approches, car les pourcentages de reconnaissance ne croient pas avec le score moyen des sons d'eau. En revanche, il est important de remarquer que les trois sons d'eau les mieux perçus sont aussi les plus reconnus. En ce sens, il peut être intéressant d'étudier les données en constituant deux groupes de sons d'eau. Le premier, le groupe $1$, est constitué des sons \textit{rain}, \textit{stream} et \textit{waves}, tandis que le groupe $2$ est constitué des sons \textit{fountain} et \textit{waterfall}. La figure \ref{groups_results} ci-dessous montre les scores obtenus pour les deux groupes (seules les mixtures générées avec la source "ventil1" ont été retenues) : 

\begin{figure}[h!]
    \centering
    \includegraphics[scale = 0.7]{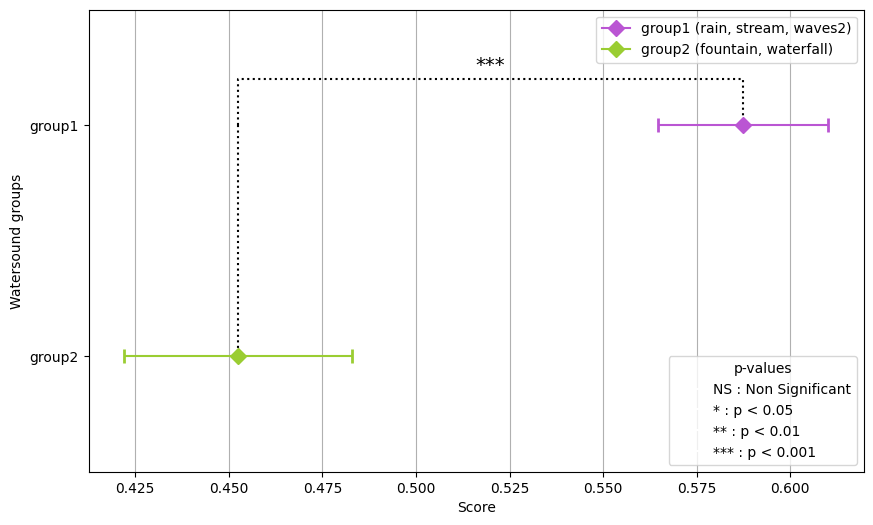}
    \caption{Scores moyens des mixtures en fonction des groupes de positifs pour la source "ventil1"}
    \label{groups_results}
\end{figure}
\newpage

Une différence significative a été relevée entre les scores moyens des deux groupes, ce qui confirme l'effet significatif des positifs sur le score. Ces résultats peuvent laisser penser que la reconnaissance des sources pour les sons d'eau a également un effet sur le score.

\subsubsection{Enquête sur les sons désirables et indésirables dans les bureaux ouverts}
\begin{figure}[h!]
    \centering
    \includegraphics[scale = 0.7]{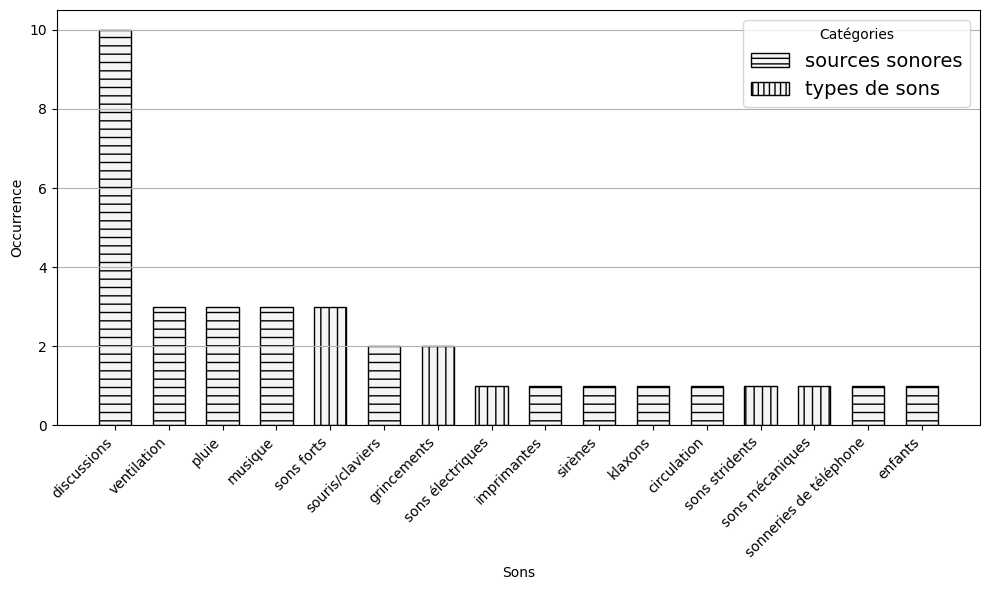}
    \caption{Occurrence des mots cités pour décrire les sons perçus comme gênants dans les bureaux ouverts}
    \label{annoying_sounds}
\end{figure}

Le questionnaire proposé à la fin de l'expérience a finalement révélé des informations sur les sons rencontrés par les participants sur leur lieu de travail. En particulier, il a été demandé aux participants d'indiquer les sons perçus comme gênants, ainsi que les sons désirés, c'est-à-dire ceux qu'ils aimeraient entendre dans un contexte de travail quotidien. La figure \ref{annoying_sounds} présente les mots cités pour décrire les sons perçus comme gênants par les participants. Ces mots sont séparés en deux catégories : ceux qui décrivent une source sonore, et ceux qui décrivent plutôt un ensemble de sons avec une caractéristique spécifique.

\newpage

Parmi les sons gênants, des sources sonores spécifiques mais aussi des types de sons sont cités. Par exemple, le bruit des discussions est le plus souvent évoqué. Certaines caractéristiques sont aussi souvent considérées comme gênantes. C'est notamment le cas des sons stridents, des sons forts ou des sons mécaniques. Par ailleurs, il est possible de distinguer les bruits intérieurs, comme les bruits d'imprimantes et la ventilation, des bruits extérieurs comme les sirènes, les klaxons et le bruit des enfants. \\ 

\begin{figure}[h!]
    \centering
    \includegraphics[scale = 0.7]{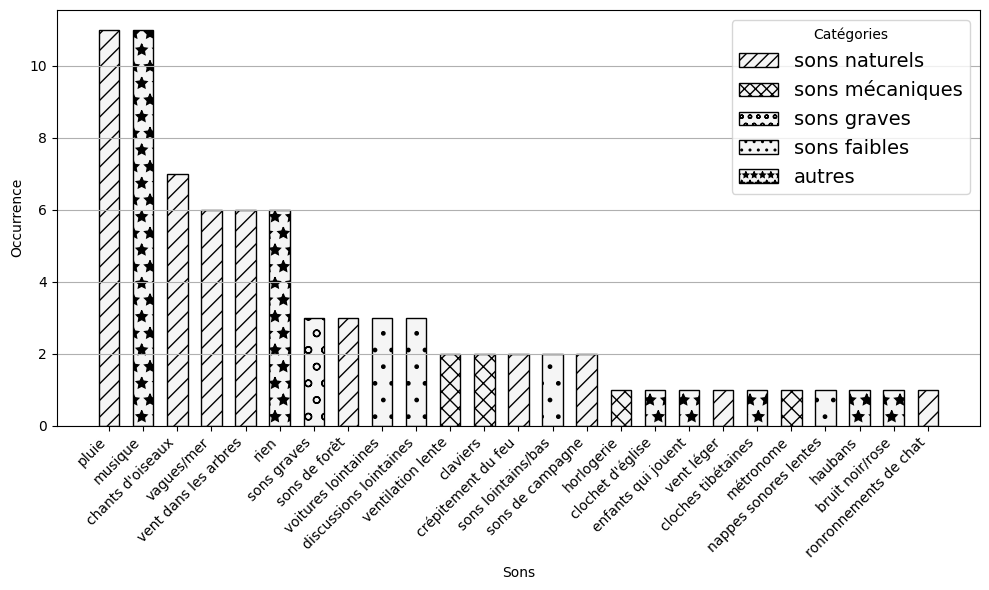}
    \caption{Occurrence des mots cités pour décrire les sons désirés dans les bureaux ouverts}
    \label{pleasant_sounds}
\end{figure}

Concernant les sons désirés, c'est-à-dire les sons que les participants aimeraient entendre dans leur environnement de travail, plus d'une vingtaine de sources sonores ont été proposées, dont certaines peuvent se regrouper dans des catégories, comme le montre la figure \ref{pleasant_sounds} suivante. 

\newpage

Les deux sons les plus désirés sont le son de la pluie et la musique. De plus, beaucoup de sons naturels ont été cités (les vagues, le vent dans les arbres, le chant des oiseaux ou encore le crépitement du feu). Certains sons mécaniques semblent aussi susciter l'agrément de certains participants, comme les bruits d'horlogerie, de claviers ou de métronome. Certains sons se retrouvent également dans leur aspect grave, comme pour la ventilation, ou faible en termes de niveau, comme pour les discussions ou les voitures "lointaines". Le nombre de mots cités relate bien du caractère subjectif de l'agrément, et il est important de noter que certains participants ont décrit les sons qu'ils aimeraient entendre avec beaucoup de précision. Par exemple, le mot 'musique', qui est apparu chez $11$ participants, était toujours accompagné d'adjectifs, comme "personnelle", "calme", "douce", "pas trop fort", "sons d'ambiance", "sans paroles" ou encore "prévisible". \\

Finalement, il est intéressant de remarquer que certains sons se trouvent à la fois dans les sons gênants et dans les sons désirés. Les bruits de claviers, des voix et des discussions environnantes, ainsi que des enfants qui jouent sont perçus très différemment. Par ailleurs, deux des sons d'eau utilisés pour l'expérience ont été cités dans les sons désirés, à savoir le son de pluie et celui de vagues. Cela confirme en partie leur statut de "positifs", car certains participants précisent que les sons d'eau ne pas sont agréables pour eux de manière générale.

\subsection{Discussion}
Le but de cette expérience est d'évaluer l'impact des approches \textit{masker} et \textit{concealer} sur l'agrément perçu, lorsque celles-ci sont utilisées sur des bruits de ventilation avec l'aide de différents sons d'eau supposés acceptés. Les résultats obtenus ne sont pas ceux espérés, car ils montrent que l'approche \textit{concealer} ne permet pas de générer des mixtures mieux perçues que celles obtenus avec l'approche \textit{masker} en termes d'agrément. \\

D'autre part, l'étude du niveau grâce au paramètre $\Delta L_{Aeq}$ semble suggérer que l'ajout des sons d'eau ne permet pas d'améliorer l'agrément perçu des deux sons de ventilation, et ce pour les deux approches utilisées. Ces résultats diffèrent de ceux obtenus par Cai et coll. pour l'approche \textit{masker}, qui ont trouvé une relation quadratique entre niveau (représenté par le paramètre \textit{MSNR}) et réduction de la gêne perçue \cite{cai_effect_2019}. Cette différence peut être due à plusieurs éléments, notamment le type et le niveau de la source choisie (un bruit de soudeuse électrique à $80dB(A)$ pour l'étude de Cai et coll., contre un bruit de ventilation à $50dB(A)$ ici), mais aussi le paramètre étudié (la gêne et l'agrément perçus sont deux paramètres bien distincts). Dans le cadre de cette étude, l'ajout de complexité à la source de bruit n'est pas désiré, du moins en termes d'agrément. \\

En revanche, au vu de l'effet du type de source et du type de positif, il serait intéressant de construire le signal \textit{concealer} avec d'autres sources et d'autres positifs, car les scores en termes d'agrément semblent très dépendants de ces paramètres pour les deux approches. Les résultats obtenus lors de cette étude peuvent par ailleurs suggérer que le cadre choisi n'est pas assez représentatif pour voir les effets des traitements proposés. Il serait par exemple possible de traiter des sources beaucoup plus gênantes et désagréables, et d'étudier les sons désirables dans les bureaux ouverts pour mieux choisir les positifs. C'est pourquoi il est important de mieux comprendre le ressenti des participants quant aux stimuli présentés. D'une part, les discussions post-expérience ont révélé que les sons ont plus souvent été évalués en termes de désagrément qu'en termes d'agrément. Beaucoup de participants ont en effet précisé qu'aucun des sons ne leur paraissait agréable, et que le choix du "meilleur" stimulus en termes d'agrément correspondait en fait au moins pire. Le statut de "positifs" de sons d'eau a été remis en cause dans le contexte étudié, car ceux-ci ont souvent été perçus comme trop "forts", trop "proches", ou encore non cohérents avec un contexte de travail, certains participants mentionnant même une sensation d'inconfort à l'écoute des stimuli. Cela ne semble cependant pas unanime, puisque quelques participants ont affirmé être sortis relaxés de l'expérience. D'autre part, les participants ont identifié certaines caractéristiques des sons comme des facteurs responsables de l'agrément ou du désagrément perçu. La sonie est restée le premier facteur responsable du désagrément, ce qui rejoint la littérature et les résultats présentés plus haut. De manière générale, les participants ont dit préférer les sons graves et homogènes. En parlant plus précisément de la création de mixtures grâce aux sons de ventilation et d'eau, celles-ci ont été perçues plus agréables lorsque le son d'eau était dans la même gamme de fréquences que le bruit de ventilation. Un "équilibre" entre les sons a aussi été mentionné pour caractériser le fait que la mixture entre deux sons est la plus agréable lorsque ceux-ci sont perçus au même niveau. \\

Finalement, l'aspect subjectif de la notion d'agrément peut imposer un choix personnalisé des positifs. Cette idée de personnalisation est d'ailleurs confirmée par l'un des participants qui mentionne qu'il préfère contrôler son environnement sonore plutôt que de laisser place à l'imprévu. L'idée d'imposer des sources sonores spécifiques, même si elles sont communément acceptées, mérite d'être plus amplement étudiée pour s'assurer de son efficacité pour un traitement visant à réduire perceptivement le bruit. Des études supplémentaires sont nécessaires pour espérer identifier des typologies de sons agréables et désagréables, et ainsi trouver de bons couples source-positif pour appliquer l'approche \textit{concealer} efficacement en termes d'agrément.

\newpage
\section{Conclusion}
Afin de proposer une solution de traitement perceptif du bruit dans les environnements intérieurs, une nouvelle approche nommée \textit{concealer} a été développée. La particularité de cette approche est la construction d'un signal \textit{concealer}, qui une fois ajouté à la source de bruit agit comme "maquilleur" et permet de créer une mixture qui soit mieux perçue que le bruit initial. Pour s'assurer de cela, le signal \textit{concealer} dépend d'un son dit positif, mais aussi de la source de bruit elle-même. Cette méthode de construction du maquilleur diffère de celle utilisée pour l'approche \textit{masker}, qui consiste simplement à définir un son positif comme maquilleur. L'idée portée par l'approche \textit{concealer} est alors de compléter la source de bruit tout en minimisant l'augmentation du niveau sonore global. 

Après avoir mis en œuvre cette approche en utilisant des bruits de ventilation comme sources et des sons d'eau comme positifs, une expérience perceptive a été menée pour tenter d'évaluer l'efficacité de cette approche face à l'approche \textit{masker} en termes d'agrément. Plusieurs hypothèses ont été émises, et l'expérience a été conçue en espérant que l'approche \textit{concealer} soit meilleure que l'approche \textit{masker}, et que l'utilisation des deux approches ait un effet positif sur l'agrément. De plus, il était attendu que l'utilisation de différents types de sons d'eau ait un impact sur l'agrément perçu. Cette dernière hypothèse s'est révélée vérifiée, car les mixtures impliquant les sons \textit{stream} et \textit{waves} ont été mieux perçus que les sons \textit{rain}, \textit{fountain} et \textit{waterfall}. En revanche, les résultats n'ont pas permis de valider l'hypothèse principale, car l'approche \textit{masker} s'est révélée meilleure que l'approche \textit{concealer}.

Par ailleurs, le rapport entre le niveau ajouté et l'agrément perçu a une tendance linéaire, c'est-à-dire que plus le niveau du son positif est élevé, plus le niveau des mixtures est élevé, et moins celles-ci sont agréables. Les mixtures générées par l'utilisation des deux approches sont donc moins bien perçues que la source de bruit mixée à un très faible de niveau de positif. 

Enfin, une enquête sur les sons jugés désirables dans un contexte de travail a révélé une grande diversité de sources sonores. La divergence des goûts des participants ne rend pas évidente l'amélioration de l'agrément en créant des mixtures composées de sources sonores spécifiques. Par extension, il serait intéressant de mener une enquête plus détaillée sur les sources sonores ou les caractéristiques des sons désirés dans le contexte des bureaux ouverts. 

Pour conclure, un examen plus approfondi de cette étude pourrait apporter des informations supplémentaires aux résultats présentés dans ce document. Par exemple, l'étude d'\textit{outliers} chez les participants pourrait conduire à des analyses partielles complémentaires. Une fois cette étude terminée, plusieurs pistes sont envisageables, notamment la réévaluation des hypothèses de travail et la redéfinition du domaine d'application de l'approche \textit{concealer}. Plutôt que de travailler sur un paramètre de niveau tel que le paramètre $\Delta L_{Aeq}$, une nouvelle idée serait de définir un cadre pour le niveau sonore de la source de bruit. Par exemple, l'approche \textit{concealer} pourrait se trouver couplée à des techniques de réduction de bruit, qui ne permettent pas de supprimer complètement le bruit ambiant. L'approche \textit{concealer} trouverait alors son utilité pour des sources avec de très bas niveaux, en tentant de modifier la perception des résidus de bruit. Les mixtures générées pourraient davantage s'apparenter aux sons positifs choisis dans ce nouveau cadre. Un tel couplage ne serait en revanche pas pertinent dans le contexte des bureaux ouverts, car cela imposerait le port d'écouteurs ou de casques munis d'une réduction du bruit. Il est alors possible de s'intéresser à d'autres terrains d'application, comme les usines dans lesquelles les travailleurs sont soumis à de hauts niveaux de bruits industriels.

\newpage
\bibliographystyle{unsrt}
\bibliography{bibliographie}

\newpage
\addcontentsline{toc}{section}{Annexes}
\section*{Annexes}

\addcontentsline{toc}{subsection}{Consigne présentée aux participants (section \ref{expe})}
\subsection*{Consigne présentée aux participants de l'expérience (section \ref{expe})}\label{consigne}

\includepdf[pages={1}]{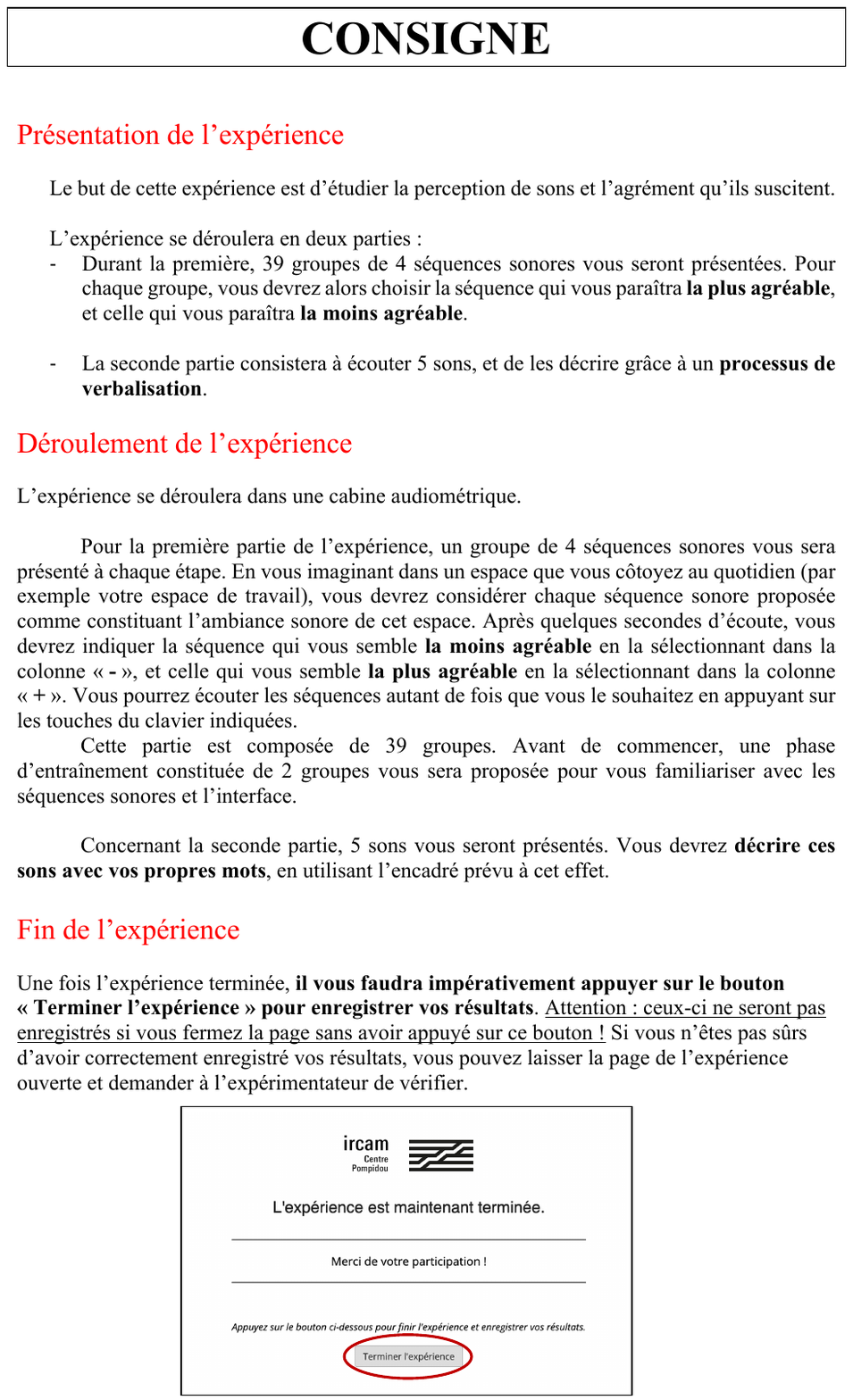}

\addcontentsline{toc}{subsection}{Feuille de consentement présentée aux participants (section \ref{expe})}
\subsection*{Feuille de consentement présentée aux participants de l'expérience (section \ref{expe})}\label{consentement}

\includepdf[pages={1}]{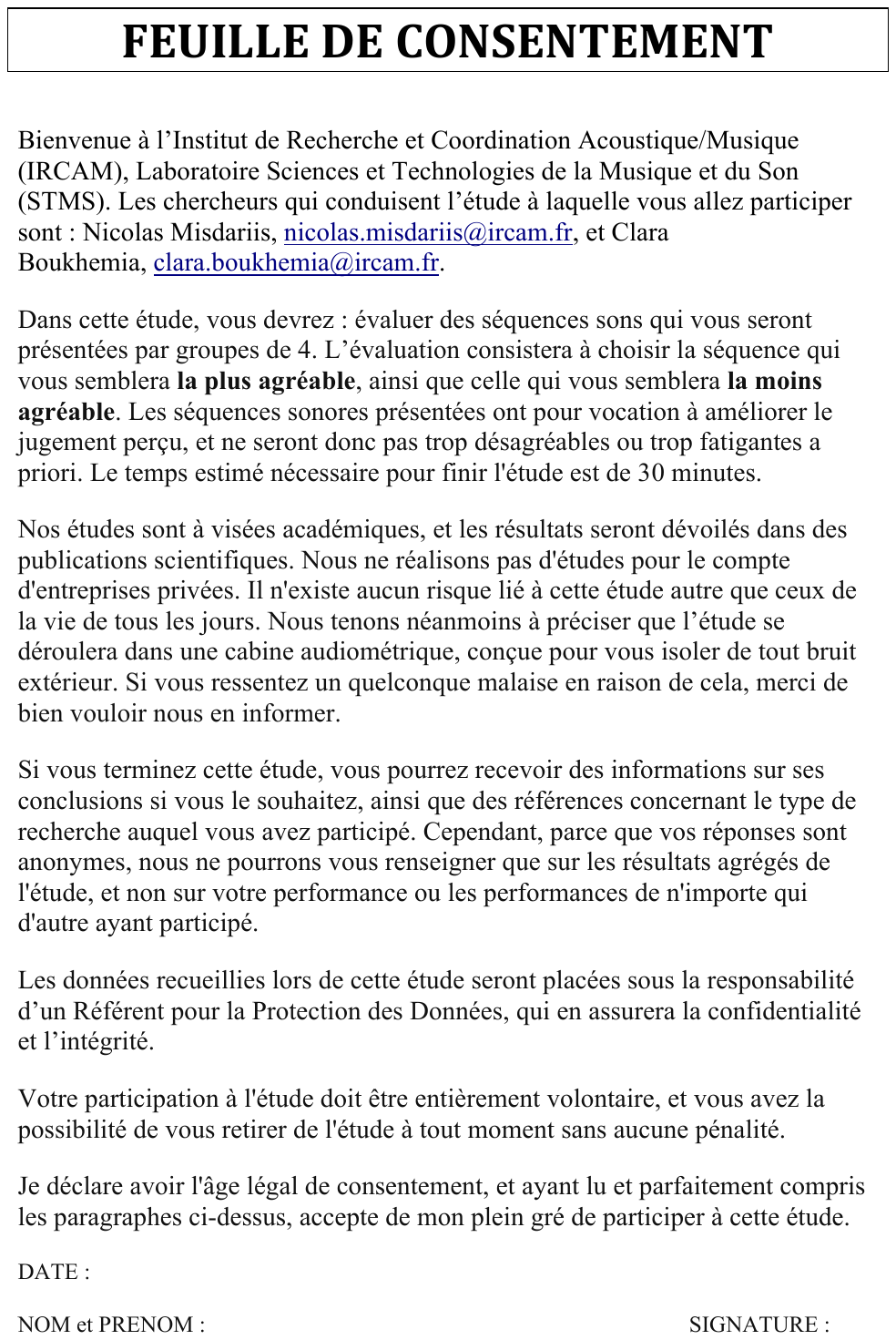}

\addcontentsline{toc}{subsection}{Questionnaire présenté aux participants à la fin de l'expérience (section \ref{expe})}
\subsection*{Questionnaire présenté aux participants à la fin de l'expérience (section \ref{expe})}\label{questionnaire}

\includepdf[pages={1-3}]{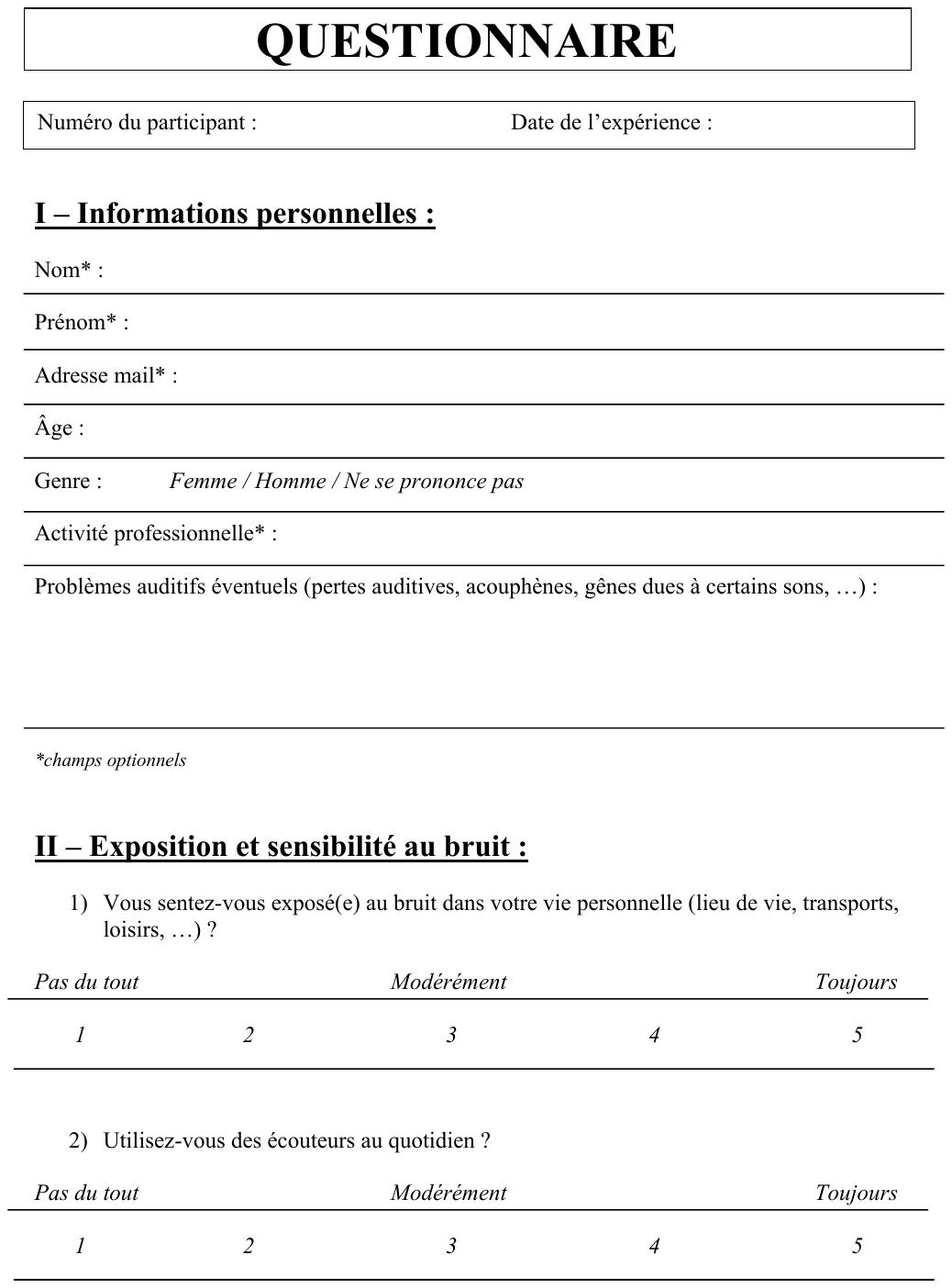}

\end{document}